\documentclass[a4paper,10pt]{article}
\pdfoutput=1 

\usepackage{jcappub} 

\usepackage{pagecolor}
\usepackage{url}
\usepackage{graphics}
\usepackage{graphicx}
\usepackage[T1]{fontenc} 
\usepackage{footnote}
\usepackage{hyperref}

\usepackage{multirow}
\usepackage{tabularx}

\usepackage{jcappub}

\def\be{\begin{equation}}
\def\ee{\end{equation}}
\def\ba{\begin{eqnarray}}
\def\ea{\end{eqnarray}}

\title{Probing primordial features with future galaxy surveys}

\author[a,b,c]{M.~Ballardini,}
\author[b,c]{F.~Finelli,}
\author[a,d,c]{C.~Fedeli}
\author[a,d,c]{and L.~Moscardini}
\affiliation[a]{DIFA, Dipartimento di Fisica e Astronomia, Alma Mater Studiorum 
Universit\`a di Bologna, Viale Berti Pichat, 6/2, I-40127 Bologna, Italy}
\affiliation[b]{INAF/IASF Bologna, via Gobetti 101, I-40129 Bologna, Italy}
\affiliation[c]{INFN, Sezione di Bologna, Via Berti Pichat 6/2, I-40127 Bologna, Italy}
\affiliation[d]{INAF/Osservatorio Astronomico di Bologna, Via Ranzani 1, I-40127 Bologna, Italy}

\emailAdd{ballardini@iasfbo.inaf.it}
\emailAdd{finelli@iasfbo.inaf.it}
\emailAdd{cosimo.fedeli@gmail.com}
\emailAdd{lauro.moscardini@unibo.it}

\abstract{
We study the capability of future measurements of the galaxy clustering power spectrum to probe 
departures from a power-law spectrum for primordial fluctuations.
On considering the information from the galaxy clustering power spectrum up to quasi-linear scales, 
i.e. $k<0.1$ h Mpc$^{-1}$, we present forecasts for DESI, Euclid and SPHEREx in combination with CMB 
measurements.
As examples of departures in the primordial power spectrum from a simple power-law, we consider four 
$Planck$ 2015 best-fits motivated by inflationary models with different breaking of the slow-roll 
approximation. 
These four representative models provide an improved fit to CMB temperature anisotropies, although 
not at statistical significant level.
As for other extensions in the matter content of the simplest $\Lambda$CDM model, the complementarity 
of the information in the resulting matter power spectrum expected from these galaxy surveys and in 
the primordial power spectrum from CMB anisotropies can be effective in constraining cosmological 
models.
We find that the three galaxy surveys can add significant information to CMB to better constrain 
the extra parameters of the four models considered.}

\begin{document}
\maketitle
\flushbottom

\section{Introduction}
\label{sec:intro}

The results from the ESA satellite $Planck$ \cite{PlanckI2013,PlanckI2015} led to important 
progresses in the context of inflation \cite{PlanckXXII2013,PlanckXX2015}.
In fact they showed how the theoretical predictions of the simplest slow-roll inflationary models, 
such as a flat Universe with nearly Gaussian adiabatic perturbations and a tilted spectrum,
provide a good fit to CMB temperature and polarization anisotropies. 
The BICEP 2/Keck Array/$Planck$ constraint on the tensor-to-scalar ratio at the scale $k_* = 0.05$ 
Mpc$^{-1}$ (the energy scale of inflation) as $r < 0.08$ ($V^{1/4} \approx 1.8 \times 10^{16}$ GeV) 
at the 95\% confidence level (CL) \cite{Ade:2015tva}, has allowed to strongly disfavour archetypal 
models such as a quadratic potential or natural inflation \cite{PlanckXX2015}.
With the most recent addition of the Keck Array 95 GHz, the constraint on primordial gravitational 
waves has been further tightened to $r < 0.07$ at 95\% CL \cite{Array:2015xqh}. 

Although a spatially flat $\Lambda$CDM model with a tilted power-law spectrum of primordial 
fluctuations provides a good fit to $Planck$ data, there are intriguing {\em features} in the 
temperature power spectrum, such as a dip at $\ell \sim 20$, a smaller average amplitude at 
$\ell \lesssim 40$ and other outliers at higher multipoles.
The features at $\ell \lesssim 40$ in the CMB temperature power spectrum generate a particular 
pattern at $k \lesssim 0.008$ Mpc$^{-1}$\footnote{Note that $k \sim 0.002$ Mpc$^{-1}$ roughly 
corresponds to $\ell \sim 20$.}, as also shown consistently by three different methods used to 
reconstruct the primordial power spectrum (PPS) of curvature perturbations with $Planck$ data 
\cite{PlanckXX2015}. Note however that none of these puzzling features in the $Planck$ temperature 
power spectrum constitute statistically significant departures from a simple power-law spectrum 
generated within the simplest slow-roll inflationary models.

There are several theoretically well motivated mechanisms during inflation which support deviations 
from a simple power law for primordial fluctuations providing a better fit to the CMB temperature 
power spectrum.
Some of these mechanisms are based on a temporary violation of the slow-roll regime for the inflaton 
field and include punctuated inflation \cite{Jain:2008dw}, a short inflationary stage preceded by a 
kinetic stage \cite{Contaldi:2003zv} or by a bounce from a contracting stage \cite{Piao:2003zm}, 
a string theory-motivated climbing phase prior to inflation \cite{Dudas:2012vv}, a sharp edge in 
the first derivative of the inflaton potential \cite{Starobinsky:1992ts}, a step in the inflaton 
potential \cite{Adams:2001vc,Hamann:2007pa}, a variation in the effective speed of sound 
\cite{Achucarro:2010da,Bartolo:2013exa,Achucarro:2012fd}, or a burst of particle production during 
inflation \cite{Barnaby:2009mc,Chantavat:2010vt}. Resonant models instead include periodic oscillations 
in the potential and therefore super-imposed periodic features to the PPS \cite{Chen:2008wn} (see \cite{Chluba:2015bqa} for a review on primordial features).
The case of periodic oscillations in axion monodromy inflation \cite{Silverstein:2008sg,McAllister:2008hb} fall in this broad class of models \cite{Flauger:2009ab}.
These features in the power spectrum are accompanied by specific templates in the bispectrum (see 
\cite{Chen:2010xka} for a review): therefore primordial features can also be searched in the 
bispectrum \cite{PlanckXVII2015} or jointly in the power spectrum and bispectrum 
\cite{Fergusson:2014hya,Fergusson:2014tza,Meerburg:2015owa}.
At present, no inflationary model fitting these features has been found to be preferred at a 
statistical significant level over more standard models \cite{PlanckXX2015,PlanckXVII2015}. 

Thanks to the sharpness of the CMB polarization transfer functions \cite{PlanckXX2015,Chluba:2015bqa}, 
future CMB polarization data will help in providing complementary information to further test if these 
deviations from a simple power-law spectrum are statistical fluctuations or are of primordial origin.
However, some of the polarization imprints of primordial features in the E-mode power spectrum 
could be confused with cosmic variance plus noise or could be degenerate with the physics of 
reionization beyond the simplest modelling of an average optical depth \cite{mortonson}. 
For primordial features fitting the oscillations at $\ell\sim 20-40$ pattern in the CMB temperature 
power spectrum, it has been estimated that the confusion of a complex reionization phase could 
decrease the statistical significance in detecting the features due to a step in the inflaton 
potential \cite{Adams:2001vc} from 8 to 5$\sigma$ for a cosmic variance dominated CMB experiment 
\cite{mortonson}.

Beyond the handle of better measurements of CMB polarization, the current snapshot of the PPS taken 
by $Planck$ \cite{PlanckXX2015,Ade:2015xua} will be also further refined by future galaxy surveys as 
J-PAS \footnote{\url{http://www.j-pas.org/}} 
\cite{Benitez:2014ibt}, 
DESI \footnote{\url{http://desi.lbl.gov/}}
\cite{Levi:2013gra,DESI:2015}, 
Euclid \footnote{\url{http://sci.esa.int/euclid/}} 
\cite{Laureijs:2011gra,Amendola:2012ys}, 
SPHEREx \footnote{\url{http://spherex.caltech.edu/}} 
\cite{Bock:2016,Dore:2014cca}, 
LSST \footnote{\url{http://www.lsst.org/}}
\cite{Abell:2009aa}, 
SKA \footnote{\url{http://www.skatelescope.org/}} 
\cite{Maartens:2015mra} and others.
Thanks to the different sensitity of the matter power spectrum to cosmology, future galaxy surveys 
will be useful to break the degeneracy among cosmological parameters encoded in the CMB angular 
power spectra of temperature and polarization.

The main goal of this paper is to assess in a quantitative way the capability of the galaxy power 
spectrum expected from future surveys having an accurate determination of redshift to probe few 
selected examples of inflationary models with a violation of the slow-roll approximation which 
provide a fit to the $Planck$ 2015 data improving on the $\Lambda$CDM model. In particular we 
restrict ourselves to DESI, Euclid and SPHEREx as a selection of future galaxy surveys which probe 
a sufficiently large volume with an accurate determination of redshift, but with different 
characteristics (see section~\ref{sec:surveys}).

Our paper is organized as follows. After this introduction, in section~\ref{sec:models} we describe 
the four representative inflationary models which are taken as examples for a better fit to the 
$Planck$ 2015 data, compared with the baseline $\Lambda$CDM model. 
In section~\ref{sec:fisher} we review the Fisher matrix approach and we describe the CMB data and 
galaxy surveys in section~\ref{sec:surveys}. 
In section~\ref{sec:results} we presents our results and we conclude in section~\ref{sec:conclusions}, 
comparing our findings to previous studies in the literature \cite{Gibelyouetal,Huangetal}.

\section{Deviations from a simple power law for primordial fluctuations consistent with {\it Planck}}
\label{sec:models}

Primordial adiabatic fluctuations with a nearly Gaussian statistics and a smooth power spectrum 
\footnote{We define the power spectrum for a variable $X$ as 
$\mathcal{P}_X(k) \equiv k^3 |X_k|^2/(2 \pi^2)$, where $X_k$ is the Fourier transform of $X$.}
are a generic prediction of standard - i.e. with a standard kinetic term - slow-roll single field 
inflationary models with a Bunch-Davies vacuum. 
The amplitude $A_{\rm s}$, the tilt $n_{\rm s}$ and the running ${\rm d} n_{\rm s}/{\rm d} \ln k$ of 
the power spectrum for the curvature perturbation $\mathcal{P}_{\cal R}(k)$:
\be
\ln \left[ \mathcal{P}_{\cal R}(k) \right] =
\ln \left( A_\mathrm{s} \right) + (n_\mathrm{s}-1) 
\ln \left( \frac{k}{k_*}\right) + \frac{1}{2} 
\frac{\mathrm{d}\ n_\mathrm{s}}{\mathrm{d}\ln k} \ln^2 \left(\frac{k}{k_*}\right) 
+ \dots
\label{eqn:spectrum}
\ee
are connected to the Hubble parameter $H$ and the Hubble flow functions (HFF) $\epsilon_{\rm i}$ 
during inflation:
\begin{align}
A_{\rm s} & \approx  \frac{H^2_*}{8 \pi^2 \epsilon_{1 \, *} } \label{eq:as_def} \\
n_{\rm s} - 1 & \approx - 2 \epsilon_{1 \, *} - \epsilon_{2 \, *} \label{eq:ns_def} \\
\frac{{\rm d}\ n_{\rm s}}{{\rm d}\ln k} & \approx  - 2 \epsilon_{1 \, *} \epsilon_{2 \, *} 
- \epsilon_{2 \, *} \epsilon_{3 \, *} \, \,,
\end{align}
where $\approx$ denotes the lowest order in the slow-roll parameters and the pedix $*$ represents the 
value of the quantity at the time in which the pivot scale $k_*$ crosses the Hubble radius during 
inflation ($k_* = a_* H_*$).
The HFF functions are defined through an hierarchy of equations involving derivatives of the Hubble 
parameter, i.e. $\epsilon_{\rm i+1}\equiv\frac{{\rm d} \ln \epsilon_{\rm i}}{{\rm d} \ln a}$ with 
$\epsilon_0 \propto H^{-1}$.
When slow-roll holds with $\epsilon_{\rm i} \ll 1$, the running and higher terms in the expansion 
\eqref{eqn:spectrum} are suppressed - being quadratic or higher order in the slow-roll parameters 
- and therefore the PPS is well approximated by a power-law. 
The extension to non-standard kinetic term introduces an additional parameter, the inflaton sound 
speed \cite{Garriga:1999vw,Gong:2015ypa}, in general time-dependent with its own hierarchy of 
higher derivative coupled to the HFFs. 

Features and/or localized bumps in the power spectra within single field inflation can occur when the 
slow-roll approximation breaks down with $\epsilon_1$ and/or $\epsilon_2$ not small. In the following 
we consider four well known examples of temporary violation of the slow-roll approximation and the 
relative analytic approximation for the resulting curvature power spectrum. 
In this paper we restrict ourselves to an inflaton with a standard kinetic term, since this class of 
models already provide a case sufficient for our purposes and hereafter we refer to the standard PPS 
$\mathcal{P}_{{\cal R},\,0}(k)$ as defined in eq.~\eqref{eqn:spectrum} with 
$\frac{{\rm d}\ n_{\rm s}}{{\rm d}\ln k} = 0$.

\begin{table*}
\centering
\caption{Best-fit for the six standard cosmological parameters and the extra parameters obtained 
with the BOBYQA algorithm \cite{BOBYQA} keeping fix the foreground parameters around their best-fit 
value for the $\Lambda$CDM case with $Planck$ TT + lowP \cite{PlanckXX2015}. 
The six cosmological parameters are: the CDM physical density $\omega_{\rm c}$, the baryon physical 
density $\omega_{\rm b}$, the Hubble parameter $H_0$, the average optical depth $\tau$, the tilt and 
the amplitude of the PPS, $n_{\rm s}$ and $A_{\rm s}$. See section~\ref{sec:models} for details on 
the extra parameters for MI, MII, MIII, MIV.}
\vspace{3mm}
\begin{tabular}{|l|ccccc|}
\hline
Parameter & Baseline & MI & MII & MIII & MIV\\
\hline
$\omega_{\rm c}$                                & $0.1198$  & $0.1197$  & $0.1198$  & $0.1201$  & $0.1184$ \\
$\omega_{\rm b}$                                & $0.02222$ & $0.02228$ & $0.02227$ & $0.02223$ & $0.2240$ \\
$H_0 [{\rm km \, s}^{-1} \, {\rm Mpc}^{-1}]$    & $67.31$   & $67.40$   & $67.32$   & $67.18$   & $68.01$ \\
$\tau$                                          & $0.078$   & $0.085$   & $0.088$   & $0.082$   & $0.085$ \\
$n_{\rm s}$                                     & $0.9655$  & $0.9647$  & $0.9655$  & $0.9647$  & $0.9723$ \\
$\ln \left( 10^{10} A_{\rm s} \right)$          & $3.089$   & $3.103$   & $3.109$   & $3.089$   & $3.102$ \\
\hline
$\lambda_{\rm c}$                               & \dots     & $0.50$    & \dots   & \dots & \dots \\
$\log_{10} \left(k_{\rm c}\ {\rm Mpc} \right)$  & \dots     & $-3.47$   & \dots   & \dots & \dots \\
\hline
$\Delta$                                        & \dots     & \dots     & $0.089$ & \dots & \dots \\
$\log_{10} \left(k_{\rm s}\ {\rm Mpc} \right)$  & \dots     & \dots     & $-3.05$ & \dots & \dots \\
\hline
$\mathcal{A}_{\rm st}$                          & \dots     & \dots     & \dots   & $0.374$ & \dots \\
$\log_{10} \left(k_{\rm st}\ {\rm Mpc} \right)$ & \dots     & \dots     & \dots   & $-3.10$ & \dots \\
$\ln x_{\rm st}$                                & \dots     & \dots     & \dots   & $0.342$ & \dots \\
\hline
$\mathcal{A}_{\rm log}$                         & \dots     & \dots     & \dots   & \dots & $0.0278$ \\
$\log_{10} \left(\omega_{\rm log} \right)$      & \dots     & \dots     & \dots   & \dots & $1.51$ \\
$\phi_{\rm log}/(2\pi)$                         & \dots     & \dots     & \dots   & \dots & $0.634$ \\
\hline
\end{tabular}
\label{tab:bestfit}
\end{table*}

\subsection{Model 1: An exponential cut-off on large scales with variable stiffness}

As first model (hereafter MI), we analyze a power-law spectrum multiplied by an exponential cut-off, 
introduced in \cite{Contaldi:2003zv}, parametrized as:
\begin{equation}
\mathcal{P_R}(k) = \mathcal{P}_{{\cal R},\,0}(k) \left\{ 1 - \exp \left[- \left( \frac{k}{k_{\rm c}}
\right)^{\lambda_{\rm c}} \right] \right\}.
\label{eqn:cutoff}
\end{equation}
Here, eq.~\eqref{eqn:cutoff} reproduces a suppression of the curvature power spectrum at large 
scales by introducing two extra parameters: the first one, $k_{\rm c}$, selects the relevant scale 
where the deviation from the smooth curvature power spectrum starts, while the second parameter, 
$\lambda_{\rm c}$, adjusts the stiffness of the suppression.

This simple parameterization is motivated by models with a kinetic stage followed by a short 
inflationary phase in which the onset of the slow-roll phase coincides with the time when the 
largest observable scales exited the Hubble radius during inflation.
\footnote{A change in the power spectrum at large scales might also be induced by first-order 
quantum gravity corrections \cite{Kamenshchik:2013msa}.}
On these largest scales, the curvature power spectrum is then strongly suppressed due to the 
kinetic energy of the inflaton, and so the CMB angular power spectra at the lowest multipoles. 
Note that the exact derivation of the PPS obtained through a matching of an initial 
kinetic-dominated regime with a de-Sitter stage shows that the large scale suppression is connected 
to the smooth nearly scale-invariant power spectrum by oscillations \cite{Contaldi:2003zv}.
However, this exact derivation leads to a smaller improvement in $\Delta \chi^2$ with respect 
to the smooth phenomenological suppression described by eq.~\eqref{eqn:cutoff}, as 
discussed in \cite{PlanckXX2015}, and therefore we choose the latter as the first representative 
case of this paper.

\subsection{Model 2: Discontinuity in the first derivative of the potential}

As a second model (hereafter MII), we consider a transition in the first derivative of the potential, 
which leads to a localized imprint in the PPS, at the scales where the transition occurred 
\cite{Starobinsky:1992ts,Gong:2005jr}. This specific model assumes a sharp change in the slope of the inflaton potential $V(\phi)$:  
\be
\label{eqn:potstaro}
V(\phi) = 
\begin{cases}
V_0 + A_+(\phi-\phi_0)\,,  &\phi \gg \phi_0\\
V_0 + A_-(\phi-\phi_0)\,,  &\phi \ll \phi_0
\end{cases}\,.
\ee
The two different slopes of the potential lead to different asymptotic values of the curvature 
power spectrum, plus an oscillatory pattern in between. 
The curvature power spectrum can be obtained analitically under the approximation 
$|A_+ \phi| \,, |A_- \phi| \ll V_0$ \cite{Starobinsky:1992ts}: 
\be
\mathcal{P}_{\mathcal{R}} (k) = \mathcal{P}_{\mathcal{R},\,0} (k) \times \mathcal{D}(y,\Delta)\,,
\ee
with:
\begin{align}
\mathcal{D}(y,\Delta) =& 1 + \frac{9\Delta^2}{2}\left( \frac{1}{y} + \frac{1}{y^3} \right)^2 
+ \frac{3\Delta}{2}\left( 4 + 3\Delta - \frac{3\Delta}{y^4} \right)^2 \frac{1}{y^2} \cos(2y) \notag\\
&+ 3\Delta\left( 1 - \frac{1 + 3\Delta}{y^2} - \frac{3\Delta}{y^4} \right)^2 \frac{1}{y} \sin(2y)\,,
\end{align}
where $y=k/k_s$ and $\Delta=(A_+-A_-)/A_+$. Here $k_s$ is the scale of the transition.

\subsection{Model 3: Step in the inflaton potential}

We now consider a different model (hereafter MIII) with a step in the inflationary 
potential~\cite{Adams:2001vc} wich predicts localized oscillations in the power spectrum. 
In this case the parameterization for the PPS is derived from the potential:
\be
V(\phi) = \frac{1}{2}m^2\phi^2 \left[ 1 + c\tanh\left(\frac{\phi-\phi_0}{b}\right) \right]\,,
\ee
where $c$ is the height and $d$ the width of the step localized at $\phi=\phi_0$. 
This step-like feature in the inflaton potential leads to a localized oscillatory pattern with 
a negligible difference in the asymptotic amplitudes of the PPS. 
An analytic approximation for the PPS describing the step in the potential has been obtained in 
refs.~\cite{Dvorkin:2009ne,Miranda:2013wxa}:
\be
\mathcal{P}_{\mathcal{R}} (k) 
= \exp \left\{ \ln\mathcal{P}_{\mathcal{R},\,0} (k) + \mathcal{I}_0 (k) + \ln\left[1 + \mathcal{I}_1^2 (k)\right] \right\}\,,
\ee
where the first-order term is:
\be
\mathcal{I}_0 (k) = \mathcal{A}_{\rm st} W'\left(\frac{k}{k_{\rm st}}\right) \mathcal{D}\left(\frac{k}{k_{\rm st} x_{\rm st}}\right)\,,
\label{eqn:step_first_order}
\ee
and the second-order contribution is \cite{Miranda:2013wxa}:
\be
\sqrt{2}\mathcal{I}_1 (k) = \frac{\pi}{2}\left(1-n_{\rm s}\right) 
+ \mathcal{A}_{\rm st} 
X'\left(\frac{k}{k_{\rm st}}\right) \mathcal{D}\left(\frac{k}{k_{\rm st} x_{\rm st}}\right) \,,
\label{eqn:step_second_order}
\ee
where $k_{\rm st}$ is the mode corresponding to the time of the transition and $x_{\rm st}$ is 
related to the duration of the violation of slow-roll. The window functions in 
eqs.~\eqref{eqn:step_first_order} and \eqref{eqn:step_second_order} are:
\begin{align}
W(x) &= \frac{3\sin(2x)}{2x^3} - \frac{3\cos(2x)}{x^2} - \frac{3\sin(2x)}{2x} \\
X(x) &= \frac{3}{x^3} \left(\sin x - x\cos x\right)^2  \,;
\end{align}
the prime in this context denotes ${\rm d}/{\rm d} \ln x$ and the damping envelope is:
\be
\label{eqn:damping}
\mathcal{D}(x) = \frac{x}{\sinh x}\,.
\ee
We can rewrite the full power spectrum of curvature perturbation as 
\cite{Dvorkin:2009ne,Miranda:2013wxa}:
\begin{align}
\mathcal{P}_{\mathcal{R}} (k) = \exp \biggr\{ &\ln\mathcal{P}_{\mathcal{R},\,0} (k) \notag\\
&+ \frac{\mathcal{A}_{\rm st} k_{\rm st}^3}{2 k^3}\left[ \left(18\frac{k}{k_{\rm st}}-6\frac{k^3}{k_{\rm st}^3}\right)\cos \left(2\frac{k}{k_{\rm st}}\right) + \left(15\frac{k^2}{k_{\rm st}^2}-9\right)\sin \left(2\frac{k}{k_{\rm st}}\right) \right] \notag\\ 
&\quad\cdot\frac{k \cosh \left(\frac{k}{k_{\rm st} x_{\rm st}}\right)}{k_{\rm st} x_{\rm st}} \notag\\
&+\ln\biggr[1 + \frac{1}{2}\biggl( \frac{\pi}{2}(1-n_{\rm s})-\frac{3\mathcal{A}_{\rm st}k_{\rm st}^3}{k^3}     \left[\frac{k}{k_{\rm st}}\cos\left(\frac{k}{k_{\rm st}}\right)-\sin\left(\frac{k}{k_{\rm st}}\right)\right] \notag\\
&\qquad\qquad\cdot\left[3\frac{k}{k_{\rm st}}\cos\left(\frac{k}{k_{\rm st}}\right)+\left(2\frac{k^2}{k_{\rm st}^2}-3\right)\sin\left(\frac{k}{k_{\rm st}}\right)\right]\frac{k \cosh \left(\frac{k}{k_{\rm st} x_{\rm st}}\right)}{k_{\rm st} x_{\rm st}} \biggr)^2 \biggr]\biggr\}\,,
\end{align}
where ${\cal A}_{\rm st}$ tunes the amplitude of the feature.

\subsection{Model 4: Logarithmic super-imposed oscillations}

As a fourth model (hereafter MIV), we study the case of logarithmic super-imposed oscillations to 
the PPS:
\be
\mathcal{P}_{\mathcal{R}} (k) = 
\mathcal{P}_{\mathcal{R},\,0} (k)\left[1 + \mathcal{A}_{\rm log} \cos\left(\omega_{\rm log} \ln\left(\frac{k}{k_*}\right) + \phi_{\rm log}
\right)
\right] \,.
\label{eqn:log_oscil}
\ee
This pattern can be generated by different mechanisms.
Axion monodromy inflation \cite{Silverstein:2008sg} motivates periodic oscillations on a
large field inflaton potential leading to an approximated analytic PPS as 
in eq.~\eqref{eqn:log_oscil} \cite{Flauger:2009ab}.
See also \cite{Flauger:2014ana} for the most recent developments including  
drifting oscillations. 
Logarithmic super-imposed oscillations can also be generated by initial quantum states
different from Bunch-Davies \cite{Martin:2003kp}.

\subsection{Current constraints from CMB}

\begin{figure}[!ht]
\centering
\includegraphics[width=9cm]{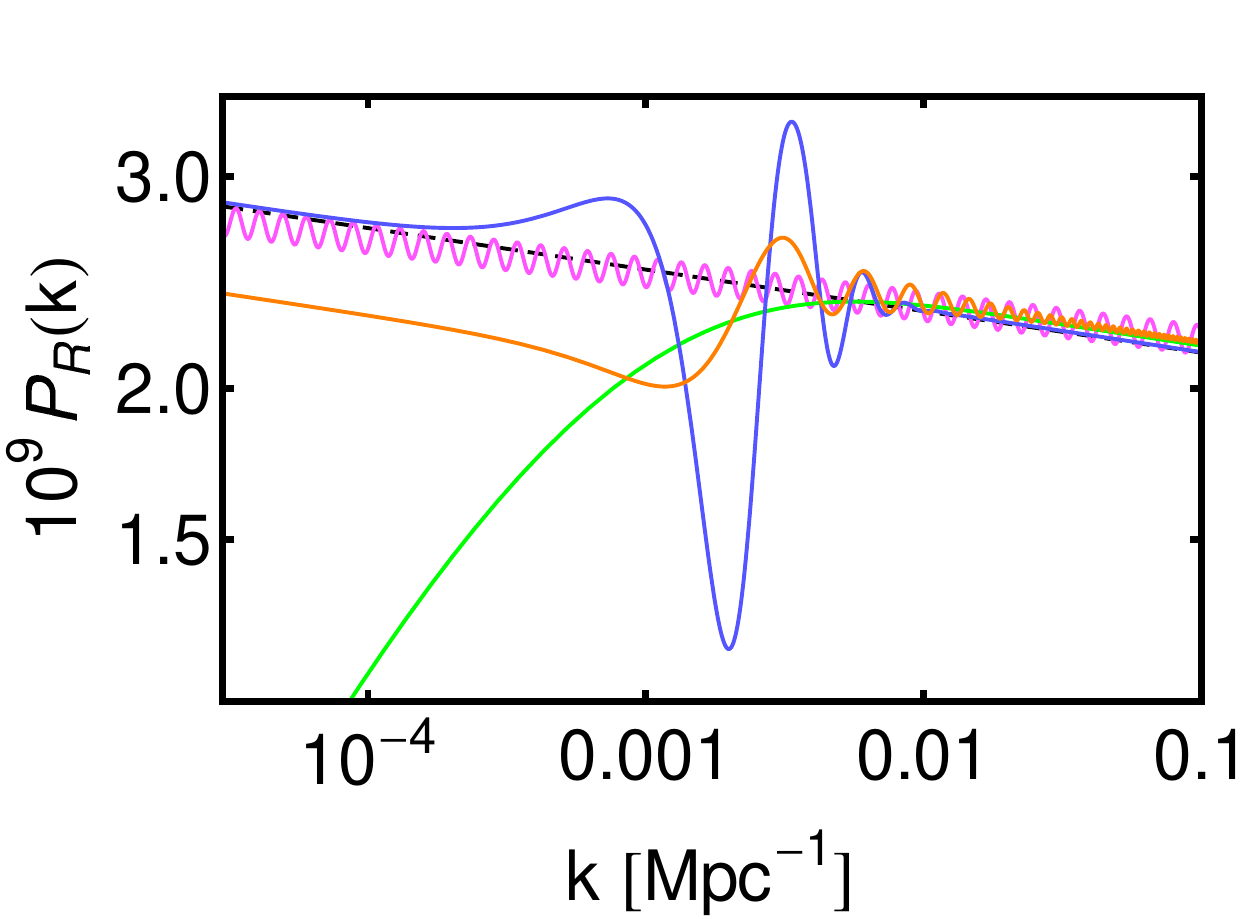}
\caption{We show the PPS for a power-law spectrum (dashed black line), for MI (green solid line), 
for MII (orange solid line), MIII (blue solid line) and MIV (magenta solid line). 
The parameters for the different models are listed in table~\ref{tab:bestfit}.\label{fig:primordial}}
\end{figure}

All the four models described above have been analysed in ref.~\cite{PlanckXX2015} 
(see also \cite{Benetti:2013cja,Miranda:2013wxa,Easther:2013kla,Chen:2014joa,
Hazra:2014goa,Hu:2014hra,Gruppuso:2015xqa,Hazra:2016fkm} for 
a non-exhaustive list of works analyzing features with $Planck$ data). 
In table~\ref{tab:bestfit} we show for each models the best-fit parameters for the standard 
cosmological parameters and for the extra parameters obtained with 
$Planck$ TT + lowP~\cite{PlanckXX2015}. 
We plot in figure~\ref{fig:primordial} the PPS for the four representative inflationary models 
and the baseline $\Lambda$CDM model. None of the four models is preferred by $Planck$ TT + lowP 
over the baseline $\Lambda$CDM model.\footnote{The Bayes factors for the four models with respect 
to the baseline $\Lambda$CDM model are $-1.4$, $-0.6$, $-0.3$, $-1.9$ \cite{PlanckXX2015}, 
respectively, with the following priors \cite{PlanckXX2015}: 
$\log_{10}\left(k_{\rm c}\ {\rm Mpc}\right) \in [-12,-3]$ and $\lambda_c \in [0,10]$ for MI, 
$\log_{10}\left(k_{\rm s}\ {\rm Mpc}\right) \in [-5,0]$ and $\Delta \in [-0.5,0.5]$ for MII, 
$\log_{10}\left(k_{\rm st}\ {\rm Mpc}\right) \in [-5,0]$, $\mathcal{A}_{\rm st} \in [0,2]$ 
and $\ln\left(x_{\rm st}\right) \in [-1,5]$ for MIII, 
$\mathcal{A}_{\rm log} \in [0,0.5]$, $\log_{10}\left( \omega_{\rm log} \right) \in [0,2.1]$ 
and $\phi_{\rm log} \in [0,2 \pi]$ for MIV.}

For the cut-off model (MI), the best-fit for the effective scale $k_c$, which marks the departure 
from a tilted power spectrum, is found at very large scales with $Planck$ 2015 data 
\cite{PlanckXX2015}, i.e. $k_{\rm c} \simeq 4 \times 10^{-4}$ Mpc$^{-1}$. The improvement in the 
fit for this model - $\Delta \chi^2 \approx -3.4$ for $Planck$ TT + lowP \cite{PlanckXX2015} - 
is due to the lower amplitude at $\ell \lesssim 40$ for the CMB temperature power spectrum.

For the other two models, which include oscillations, the effective scale of the feature is 
instead of the order of $10^{-3}$ Mpc$^{-1}$. 
For the second model (MII) the improvement in the fit - $\Delta \chi^2 \approx -4.5$ for 
$Planck$ TT + lowP~\cite{PlanckXX2015} - is due either to the lower amplitude at $\ell \lesssim 40$ 
and to the feature at $\ell \sim 20$. 
The model with a step in the potential (MIII) fits much better the feature at $\ell \sim 20$ 
and provides $\Delta \chi^2 \approx -8.6$.
Note that a low value for the quadrupole and oscillations at $\ell \sim (20 - 40)$ was also present 
in WMAP data \cite{Peiris:2003ff}; however, only the precision of the $Planck$ measurement in the 
region of the acoustic peaks has shown how the models discussed so far provide a better fit to CMB 
data than the simplest extended model with a negative running of the scalar spectral index.

The model with logarithmic oscillations (MIV) provides $\Delta \chi^2 \approx -10.8$ for 
$Planck$ TT + lowP~\cite{PlanckXX2015}, which is mainly driven by fitting outliers from the best-fit 
$\Lambda$CDM at multipoles $\ell \gtrsim 100$.   

\section{Combined Forecast for CMB and LSS}
\label{sec:fisher}

We use the Fisher matrix technique \cite{Tegmark:1996bz} for our science forecasts (as in the 
forecasts of DESI \cite{Levi:2013gra}, Euclid \cite{Amendola:2012ys} 
and SPHEREx \cite{Dore:2014cca}). The Fisher matrix technique approximates the logarithm of the likelihood as 
a multivariate Gaussian in the cosmological parameters $\{ \theta_i \}$ around a maximum at $\{\bar{\theta}_i \}$, which 
is a sufficient approximation for our purposes (see \cite{Perotto:2006rj,Wolz:2012sr} for a 
comparison of Fisher matrix approach with a full likelihood Monte Carlo Markov Chain 
for cosmological models including massive neutrinos and dynamical dark energy, respectively).
The logarithm of the likelihood can be expanded as a Taylor series and the Fisher matrix can 
be approximated as the second derivative around the peak:
\be
F_{ij} = - \left\langle  \frac{\partial^2 \ln \mathcal{L}}{\partial \theta_i\partial\theta_j} \right\rangle
= - \frac{\partial^2 \ln \mathcal{L}}{\partial \theta_i\partial\theta_j} \bigg|_{\bar{\theta}} \,. 
\ee
The diagonal elements of the inverse Fisher matrix bound the parameter variances:
\be
\label{eqn:fisher}
\text{Cov}(\theta_i,\theta_i) \geq \left[F^{-1}\right]_{ii}\,,
\ee
where we perform the inversion of the matrix before.

In the next two subsections we describe the CMB and LSS likelihoods and the corresponding 
Fisher matrices \cite{Eisenstein:1998hr}, which will be added to obtain our combined results.

\begin{figure}[!ht]
\centering
\includegraphics[width=5 cm]{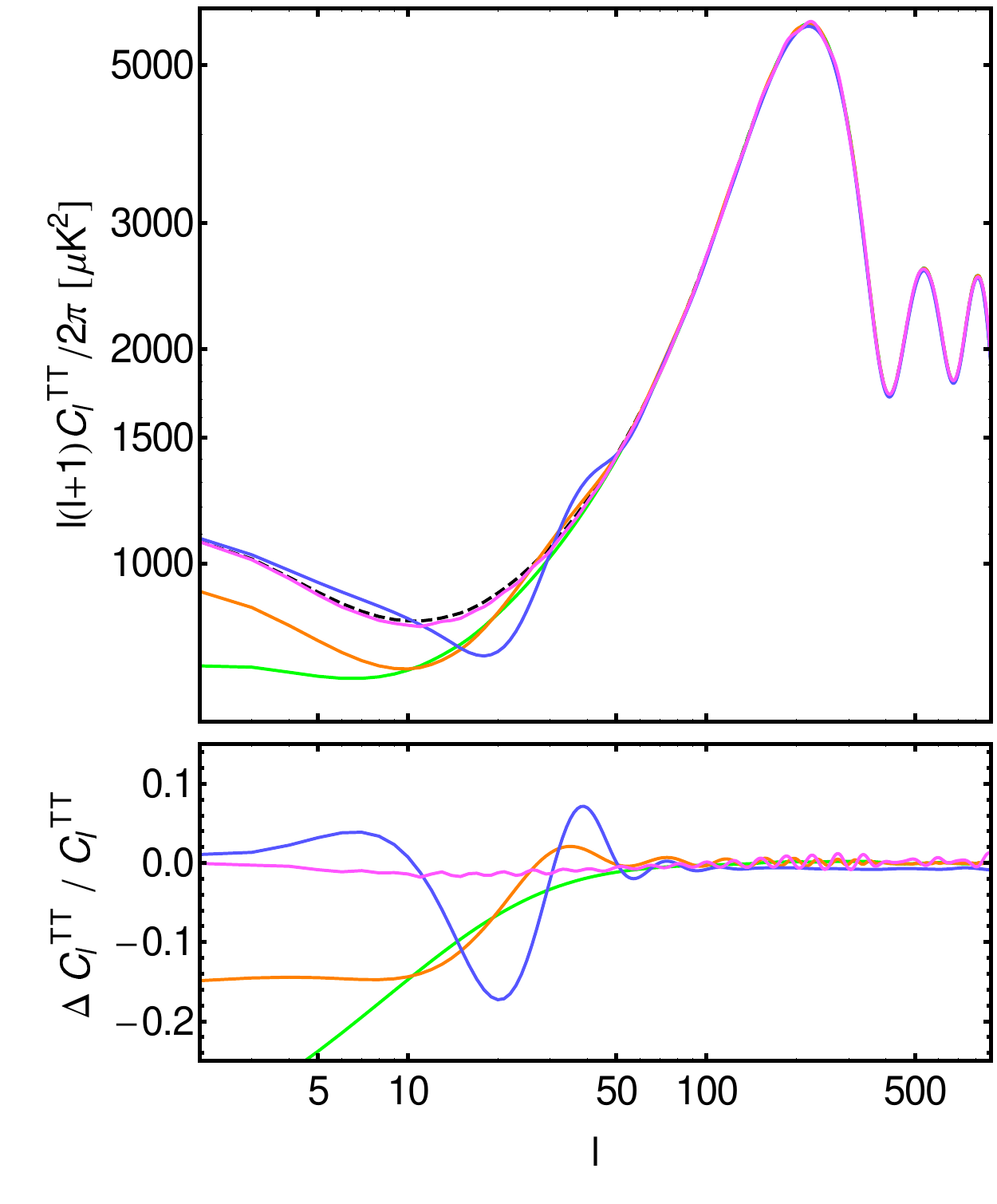}
\includegraphics[width=5 cm]{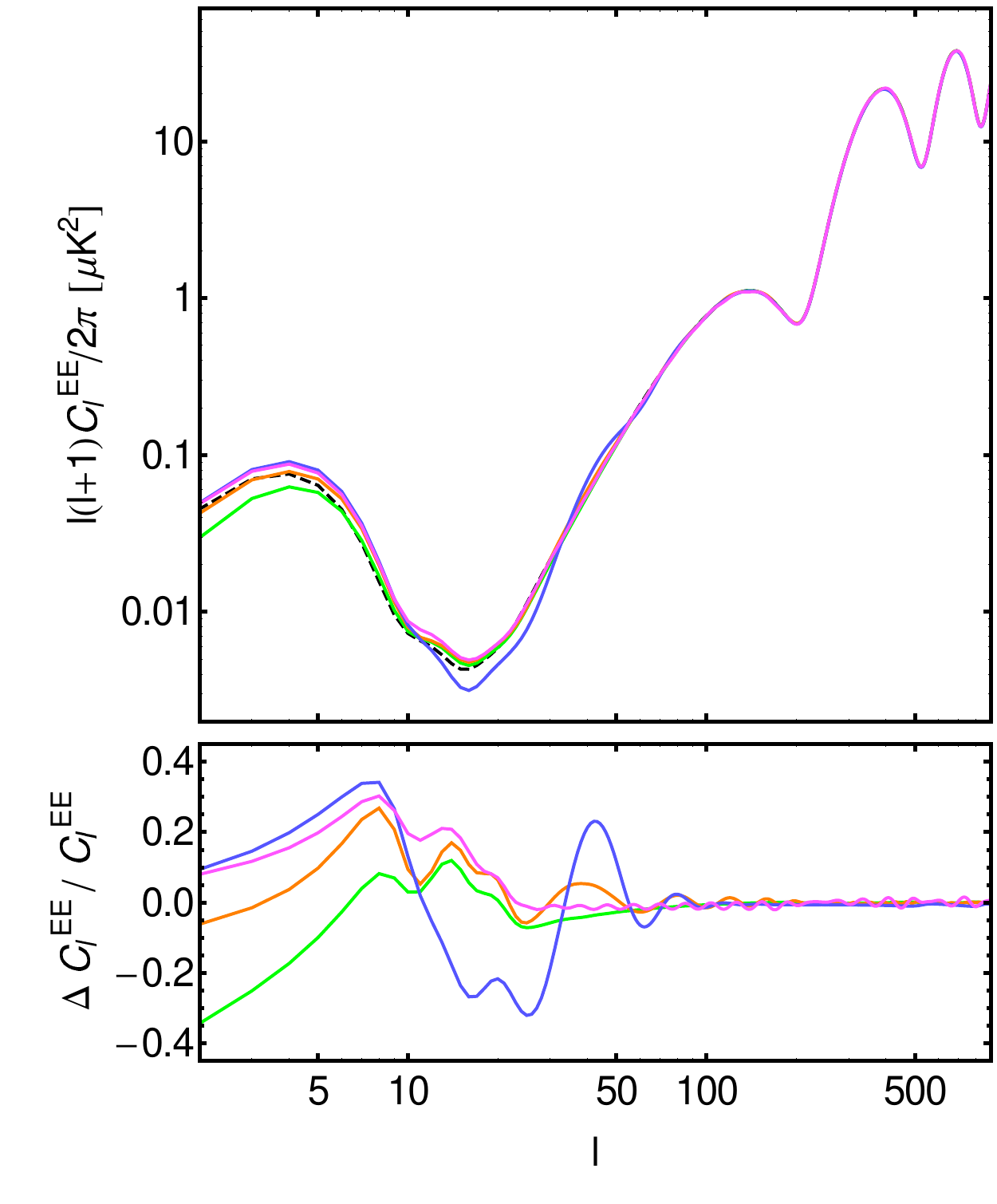}
\includegraphics[width=5 cm]{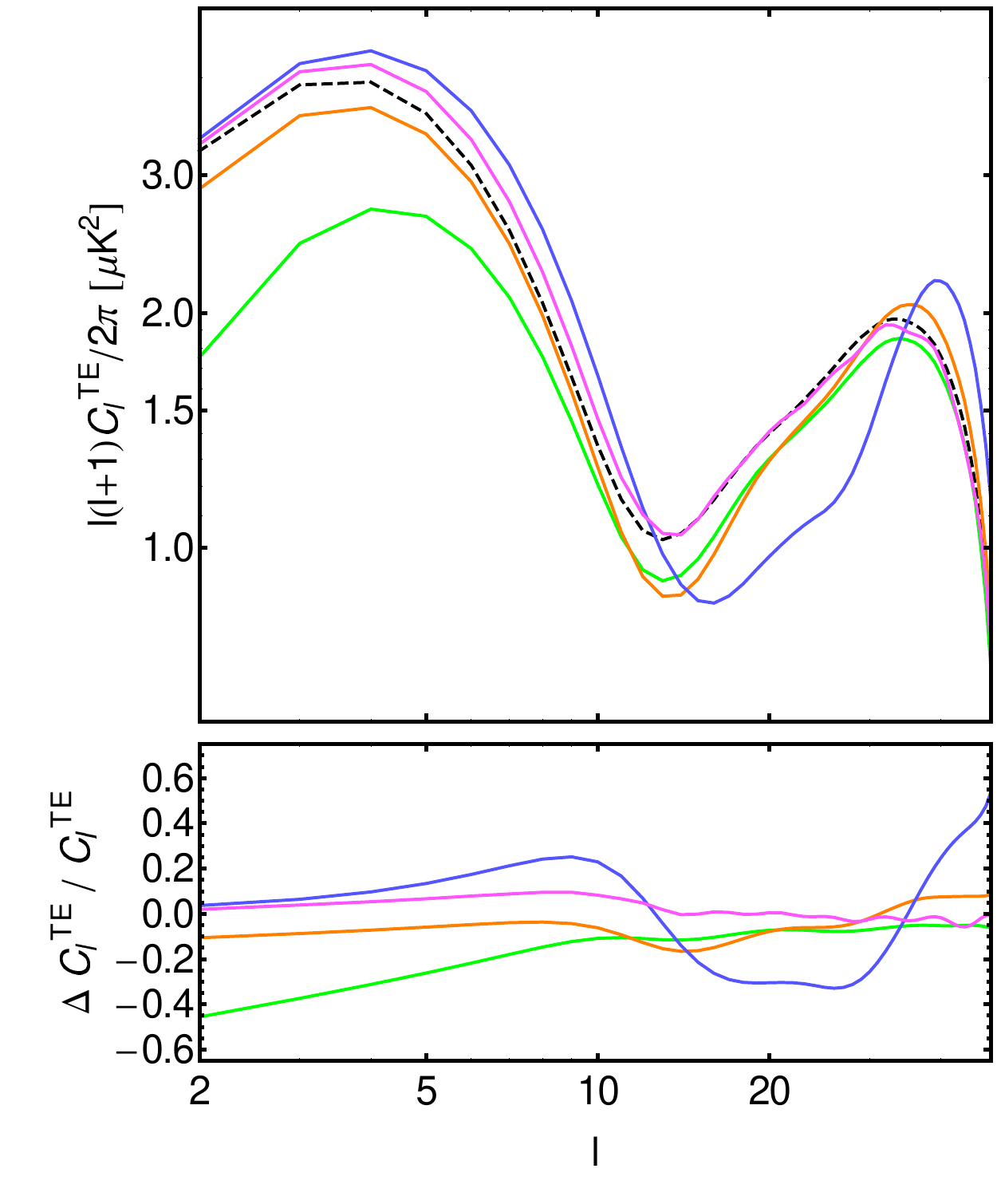}
\caption{Lensed angular power spectra TT (left panel), 
EE (middle panel), TE (right panel) for the baseline $\Lambda$CDM model (black dashed line), 
for MI (green solid line), for MII (orange solid line), MIII (blue solid line) and 
MIV (magenta solid line). In the bottom panels we display the corresponding relative differences of 
the models with respect to the baseline $\Lambda$CDM model.\label{fig:cls}}
\end{figure}

\begin{figure}[!ht]
\centering
\includegraphics[width=9 cm]{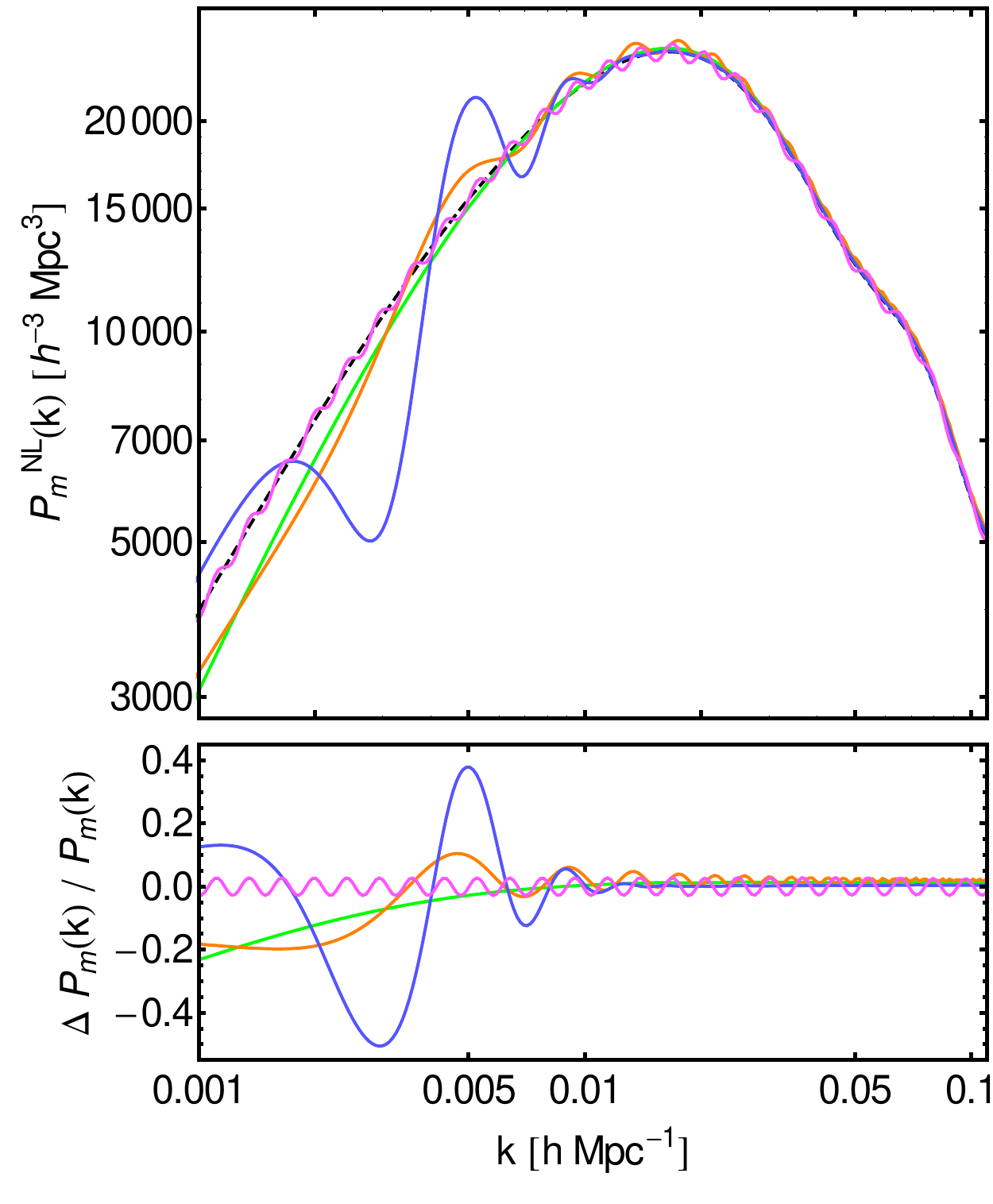}
\caption{In the top panel we show the non-linear matter power spectrum for $\Lambda$CDM 
(dashed black line), for MI (green solid line), for MII (orange solid line), MIII (blue solid line)
and MIV (magenta solid line). In the bottom panel we display the corresponding relative differences 
of the models with respect to the baseline $\Lambda$CDM model.\label{fig:matter}}
\end{figure}

\subsection{Fisher matrix for CMB}

We consider as CMB observables the lensed TT, EE and TE angular power spectra with the 
best-fit parameters showed in table~\ref{tab:bestfit}.
In figure~\ref{fig:cls} we show these angular power spectra for the four best-fits and the relative 
differences with respect to the baseline $\Lambda$CDM model.

To compute the Fisher matrix for the polarized CMB angular power spectra 
\cite{Knox:1995dq,Jungman:1995bz,Seljak:1996ti,Zaldarriaga:1996xe,Kamionkowski:1996ks} 
we use eq.~\eqref{eqn:fisher} with observables being the autocorrelators of temperature and E-mode 
polarization, and their cross-correlation.\footnote{In this paper we restrict ourselves to TT, EE, 
TE, although extensions of the models considered here exhibit a non-trivial tensor-to-scalar ratio 
if the energy-scale of inflation is sufficiently large \cite{Nicholson:2007by}.}
We consider the lensed polarized CMB angular spectra, although not taking into account the 
CMB deflection angle information as a separate observable. The covariance matrix for the observables 
is given by:
\be
\label{eqn:fisherCMB}
{\bf{\rm F}}_{ij}^\mathrm{CMB} = \sum_{\ell} \sum_{X,Y} \frac{\partial C_\ell^{X}}{\partial\theta_i}
\left(\mathrm{Cov}_\ell\right)^{-1}_{XY} \frac{\partial C_\ell^{Y}}{\partial\theta_j}\,,
\ee
where we consider $X,\,Y \in ($TT$,\,$EE$,\,$TE$)$ and the matrix $\text{Cov}_\ell$ is the symmetric 
angular power spectrum covariance matrix at the $\ell$-th multipole:
\be
\mathrm{Cov}_\ell = \frac{2}{(2 \ell+1) f_\mathrm{sky}}
\label{eqn:covmatCMB}
\begin{bmatrix}
(\bar{C}_\ell^{\rm TT})^2 & (\bar{C}_\ell^{\rm TE})^2 & \bar{C}_\ell^{\rm TT}\bar{C}_\ell^{\rm TE} \\
(\bar{C}_\ell^{\rm TE})^2 & (\bar{C}_\ell^{\rm EE})^2 & \bar{C}_\ell^{\rm EE}\bar{C}_\ell^{\rm TE} \\
\bar{C}_\ell^{\rm TT}\bar{C}_\ell^{\rm TE} & \bar{C}_\ell^{\rm EE}\bar{C}_\ell^{\rm TE} & \left((\bar{C}_\ell^{\rm TE})^2+\bar{C}_\ell^{\rm TT}\bar{C}_\ell^{\rm EE}\right)/2 
\end{bmatrix}\,,
\ee
where $\bar{C}_\ell^{X} = C^{X}_\ell + N^{X}_\ell$ is the sum of the signal and the noise, 
with $N^{\rm TE}_\ell=0$. Here $N^{X}_\ell=w^{-1}_X b^{-2}_\ell$ is the isotropic noise convolved 
with the instrument beam, $b_\ell^2$ is the Gaussian beam window function, with 
$b_\ell = e^{-\ell (\ell + 1) \theta_{\rm FWHM}^2/16\ln 2}$; $\theta_{\rm FWHM}$ is the full 
width half maximum (FWHM) of the beam in radians; $w_{\rm TT}$ and $w_{\rm EE}$ are the inverse 
square of the detector noise level on a steradian patch for temperature and polarization, 
respectively. For multiple frequency channels, $w_X b_\ell^2$ is replaced by the inverse 
noise-weighted sum over channels.
Eq.~\eqref{eqn:covmatCMB} includes sampling variance which accounts also for the loss 
of information due to partial sky coverage for $f_{\rm sky}<1$. We compute the CMB angular 
power spectra in eqs.~\ref{eqn:fisherCMB} and \ref{eqn:covmatCMB} using a modified version of 
the publicly available Einstein-Boltzmann code CAMB 
\footnote{\url{http://camb.info/}} \cite{Lewis:1999bs,Howlett:2012mh} in order to calculate $C_\ell$ at each multipoles.

We consider a ${\rm F}_{ij}^{\rm CMB}$ which represents the CMB measurements at the 
timescales of the galaxy surveys analyzed here. In this paper we restrict ourselves to noise 
sensitivity and angular resolution to characterize the uncertainties in the CMB temperature and 
polarization spectra, although we know that the accuracy of CMB anisotropies measurements are not 
governed only by noise sensitivity and angular resolution, but limited in temperature at high 
multipoles by foreground residuals/secondary anisotropies and at low multipoles in polarization by 
the Galactic emission.
Since the time scales of the surveys are different, and not only the $Planck$ final data in 
temperature and polarization, but possibly other measurements of CMB E-mode polarization on a 
large fraction of the sky, such as from AdvACTpol~\cite{Henderson:2015nzj}, 
CLASS~\cite{Essinger-Hileman:2014pja}, LSPE~\cite{:2012goa}, will be available, we consider two 
settings, one more conservative (hereafter CMB-1) and another with better sensitivity and angular 
resolution (CMB-2).

We therefore consider the $Planck$ 143 GHz channel as CMB-1 and the inverse noise weighted 
combination of the $Planck$ 70, 100, 143 and 217 GHz channels as CMB-2. 
We consider updated full mission sensitivities and angular resolution as given in \cite{PlanckI2015}.
We consider $f_{\rm sky} = 0.75$, a sum up to $\ell_{\rm max} = 2500$ in eq.~\eqref{eqn:fisherCMB}.

\subsection{Fisher matrix for spectroscopic galaxy surveys}

We consider the galaxy clustering as observable for the Fisher LSS forecast in 
eq.~\eqref{eqn:fisher}. 
The simplest model for the observed galaxy power spectrum assumes a linear and 
scale-independent galaxy bias, with redshift space distorsions due to small peculiar velocities not 
associated to the Hubble flow \cite{Kaiser:1987qv} given by:
\be
P_{\rm g}({\bf k},z) = b(z)^2\left[1+\beta(k,z)\mu^2\right]^2 P_{\rm m}(k,z)
e^{-k^2\mu^2\sigma_{\rm tot}^2} \,,
\ee
where $b$ is the bias, which maps the mass field into the galaxy density one, $\beta \equiv f/b$ 
with $f\equiv {\rm d} \ln D/{\rm d}\ln a$ is the growth rate, $\mu$ is the angle to the line of sight 
and $P_{\rm m}$ represents the dark matter power spectrum in real space. 
In redshift space, the observed galaxy power spectrum can be modelled by including inaccuracies in 
the observed redshifts as \cite{White:2008jy} in addition to the linear Kaiser effect. 
We define $\sigma_{\rm tot}$ as $\sqrt{\sigma_{\rm v}^2+\sigma_{\rm r}^2}$ where
$\sigma_{\rm {\bf r}} \simeq \sigma(z)c/H(z)$ is the spectrometric redshift error. We 
parametrized the redshift dependence as $\sigma(z) = \bar{\sigma}_z(1+z)$ where $\bar{\sigma}_z$ 
is the average redshift error within a redshift bin. Moreover, $\sigma_{\rm v}^2$ is the square 
of the velocity dispersion, which depends from the velocity power spectrum, and we choose a value 
of $\sigma_{\rm v}=7$ Mpc for our fiducial value 
\cite{Li:2007rpa,Bull:2014rha}, which corresponds to a velocity dispersion of $\sim 500$ km/s.

For a Poisson sampled density field, we obtain in addition a constant shot-noise contribution to 
the power due to the finite number of galaxies per bin $P_\textup{shot}(z)$:
\be
P_{\rm obs}({\bf k},z) = P_{\rm g}({\bf k},z) + P_\textup{shot}(z)\,.
\ee

We include in the observed galaxy power spectrum the geometrical effects due to the incorrect 
assumption of the reference cosmology with respect the true/fiducial one \cite{Seo:2003pu,Song:2008qt}:
\be
\label{eqn:obsPdk}
\widetilde{P}_{\rm obs}(k^{\rm ref}, \mu_{\rm k}^{\rm ref},z) = 
\left( \frac{D^{\rm ref}_{\rm A}(z)}{D_{\rm A}(z)} \right)^2
\frac{H(z)}{H^{\rm ref}(z)} P_{\rm g}(k^{\rm ref}, \mu_{\rm k}^{\rm ref},z) 
+ P_\textup{shot}(z)\,,
\ee
where the prefactor is the Alcock-Paczynski (AP) effect \cite{Alcock:1979mp,Seo:2003pu}. 
The true wave-numbers $k$ and the direction cosine $\mu_{\rm k}$ calculated by assuming the reference 
cosmology are related to the ones in the true cosmology through:
\be
\label{eqn:kref}
k = k^{\rm ref}\,\sqrt{\left(\frac{H\mu_{\rm k}^{\rm ref}}{H^{\rm ref}}\right)^2
+\left(\frac{D_{\rm A}^{\rm ref}}{D_{\rm A}}\right)^2\left[(\mu_{\rm k}^{\rm ref})^2-1\right]} \,,
\ee 
and 
\be
\label{eqn:muref}
\mu_{\rm k} = \mu_{\rm k}^{\rm ref}\left(\frac{H}{H^{\rm ref}}\right)^2\frac{k^{\rm ref}}{k} \,.
\ee

Under the assumption that the density field has a Gaussian statistics and uncorrelated Fourier modes, 
the Fisher matrix for the broadband power spectrum eq.~\eqref{eqn:obsPdk}, for a given redshift bin 
with $\bar{z}$ as centroid value, is \cite{Tegmark:1997rp}:
\begin{align}
\label{eqn:fisherLSS}
{\bf{\rm F}}_{ij}^\textup{gg}(\bar{z}) &= 
\int^{k_{\rm max}}_{k_{\rm min}} \frac{d^3{\bf k}}{2(2\pi)^3}
\frac{\partial \ln \widetilde{P}_{\rm obs}({\bf k}|\bar{z})}{\partial \theta_i}\Big|_{\bar{\theta}}
\frac{\partial \ln \widetilde{P}_{\rm obs}({\bf k}|\bar{z})}{\partial \theta_j}\Big|_{\bar{\theta}}
V_\textup{eff}({\bf k}|\bar{z}) \\
&=\int^{k_{\rm max}}_{k_{\rm min}} \frac{k^2d k}{(2\pi)^2} \int^{1}_{0}d\mu
\frac{\partial \ln \widetilde{P}_{\rm obs}(k,\mu|\bar{z})}{\partial \theta_i}\Big|_{\bar{\theta}}
\frac{\partial \ln \widetilde{P}_{\rm obs}(k,\mu|\bar{z})}{\partial \theta_j}\Big|_{\bar{\theta}}
V_\textup{eff}(k,\mu|\bar{z}) \,,
\end{align}
where the effective volume of the survey in Fourier space, which determines the mode counts, 
is \cite{Feldman:1993ky}:
\begin{align}
V_\textup{eff}(k,\mu|\bar{z}) &= \int^{k_{\rm max}}_{k_{\rm min}} \frac{d^3{\bf r}}{(2\pi)^3} 
\left[\frac{\bar{n}_{\rm g}(\bar{z})\widetilde{P}_{\rm obs}(k,\mu|\bar{z})}{\bar{n}_{\rm g}(\bar{z})\widetilde{P}_{\rm obs}(k,\mu|\bar{z})+1}\right]^2 \\
&\simeq \left[\frac{\bar{n}_{\rm g}(\bar{z})\widetilde{P}_{\rm obs}(k,\mu|\bar{z})}{\bar{n}_{\rm g}(\bar{z})\widetilde{P}_{\rm obs}(k,\mu|\bar{z})+1}\right]^2 V_{\rm surv}(\bar{z})\,,
\end{align}
which depends on the geometrical volume of the survey, $V_\textup{surv}$, and on the average number 
density, $\bar{n}_{\rm g}$, of tracers in a specific redshift bin. 
We consider the information up to the quasi non-linear scales, i.e. $k \leq 0.1$ h Mpc$^{-1}$ in all 
redshift bins. In these equations $k_{\rm min}$ is set by the slice volume in the corresponding $i$-th 
redshift bin, i.e. $k_{\rm min}(\bar{z}_{\rm i})=2\pi/\sqrt[3]{V_\text{surv}(\bar{z}_{\rm i})}$. 
We consider $k \ge k_{\rm min}(\bar{z}_{\rm i})$ with a linear binning scheme, adopting the minimum 
$\Delta k =  1.4 / \sqrt[3]{V_\text{surv}(\bar{z}_{\rm i})}$ for which the correlation between 
different bins can be neglected according to ref.~\cite{Abramo:2011ph}.
Eq.~\eqref{eqn:fisherLSS} can be therefore rewritten as a binned sum over $k$ and $\mu$:
\be
\label{eqn:fisherLSS2}
{\bf{\rm F}}_{ij}^{gg}(\bar{z}) = \sum_{k,\mu} 
\frac{\partial \ln \widetilde{P}_{\rm obs}(k,\mu|\bar{z})}{\partial \theta_i}\Big|_{\bar{\theta}}
\left[\text{Cov}_{\bf k}(\bar{z})\right]^{-1}
\frac{\partial \ln \widetilde{P}_{\rm obs}(k,\mu|\bar{z})}{\partial \theta_j}\Big|_{\bar{\theta}}\,,
\ee
where
\be
\text{Cov}_{\bf k}(\bar{z}) = 
\frac{(2\pi)^2}{k^2 \Delta k\Delta \mu}\frac{1}{V_{\rm eff}(k,\mu|\bar{z})}\,.
\ee
We consider 10 bins in $\mu$ between 0 and 1. The derivative in eq.~\eqref{eqn:fisherLSS2} is 
\cite{White:2008jy,Samushia:2010ki}:
\begin{align}
\frac{d \ln P_{\rm g}}{d \theta_i}(k,\mu_{\rm k}|z_{\rm i}) \simeq 
&\ \frac{\partial \ln P_m(k|z_{\rm i})}{\partial \theta_i} 
+\frac{2\mu_{\rm k}^2}{1+\beta(k|z_{\rm i})\mu_{\rm k}^2}\frac{\partial \beta(k|z_{\rm i})}{\partial \theta_i} \notag\\
&+\left[1+\frac{4\beta(k|z_{\rm i})\mu_{\rm k}^2}{1+\beta(k|z_{\rm i})\mu_{\rm k}^2}\left(1-\mu_{\rm k}^2\right)
+\mu_{\rm k}^2 \frac{\partial \ln P_m(k|z_{\rm i})}{\partial \ln k} \right]
\frac{\partial \ln H(z_{\rm i})}{\partial \theta_i} \notag\\
&+\left[-2+\frac{4\beta(k|z_{\rm i})\mu_{\rm k}^2}{1+\beta(k|z_{\rm i})\mu_{\rm k}^2}\left(1-\mu_{\rm k}^2\right)
-\left(1-\mu_{\rm k}^2\right) \frac{\partial \ln P_m(k|z_{\rm i})}{\partial \ln k} \right]
\frac{\partial \ln D_{\rm A}(z_{\rm i})}{\partial \theta_i} \notag\\
&+ \frac{2}{1+\beta(k|z_{\rm i})\mu_{\rm k}^2}\frac{\partial \ln b(z_{\rm i})}{\partial \theta_i} 
+ \frac{1}{P_{\rm obs}(k,\mu_{\rm k}|z_{\rm i})}\frac{\partial P_{\rm shot}(z_{\rm i})}{\partial \theta_i} 
- k^2 \mu_{\rm k}^2 \frac{\partial \sigma_{\rm tot}^2}{\partial \theta_i} \,,
\end{align}

We compute the Fisher matrix using CAMB to calculate the 
exact linear matter power spectrum for each bin and Halofit \cite{Takahashi:2012em} to include 
its non-linear evolution on small scales. The derivatives in eq.~\eqref{eqn:fisherLSS2}, 
and consistently in eq.~\eqref{eqn:fisherCMB}, are calculated numerically with the symmetric 
difference quotient:
\be
\frac{\partial \ln f({\bf k}|\bar{\theta})}{\partial \theta_i} \simeq
\frac{f({\bf k}|\theta_i+\Delta_i)
-f({\bf k}|\theta_i-\Delta_i)}{2\Delta_i f({\bf k}|\bar{\theta})}\,,
\ee
where we choose the stepsize $\Delta_i$ in order to reproduce the 68\% confidence limit 
of the parameters $\theta_i$. We have checked that the results are stable with respect to 
changes in the stepsize.

We divide the array of independent parameters $\theta$ made by three subgroups:
the cosmological parameters 
$\theta_0 = \{\Omega_{\rm c},\Omega_{\rm b},h_0,\tau,n_{\rm s},\ln\left(10^{10}A_{\rm s}\right)\}$, 
the extra parameters which describe the parametrization of 
the primordial power spectrum $\theta_{\rm ext}$ and the nuisance parameters 
$\theta_{\rm nui} = \{b,P_{\rm shot},\sigma^2_{\rm tot}\}$ according to ref.~\cite{White:2008jy}. 
We consider a set of nuisance parameters per bin in order to avoide any possible prior 
information on them.
In this analysis we marginalize over $\theta_{\rm nui}$ and we assume that they do not 
depend on the cosmological parameters, $\theta_0$, and the extra ones, $\theta_{\rm ext}$.

\section{A selection of future LSS catalogs}
\label{sec:surveys}

In the coming years, an enormous effort will be put in the realization of large galaxy surveys having 
the primary goal of determining the main cosmological parameters exploiting the information hidden 
in the clustering properties. The power of a survey is based on its capability of providing the 
most accurate positions and redshifts (corresponding to distances, when a cosmological model is 
assumed) for the largest number of well-classified objects, distributed over the widest possible 
volumes. Different strategies have been designed to optimize the scientific return of a galaxy 
surveys mantaining the request in terms of observational time under control.

In this section we describe the three spectroscopic projects used in the following section for our 
forecasts. They are different examples of future LSS surveys having  a wide sky coverage: DESI is an 
example of ground-based survey following the multi-tracer approach; Euclid is a spectroscopic survey 
from space observing mostly H$\alpha$ emitting galaxies at relatively high redshifts ($0.9<z<1.8$) 
with high redshift accuracy; SPHEREx is a proposed space mission covering a very large sky area, 
with the peak of the observed galaxy density at lower redshifts compared to the other two surveys.

\begin{figure}[!ht]
\centering
\includegraphics[width=5.05 cm]{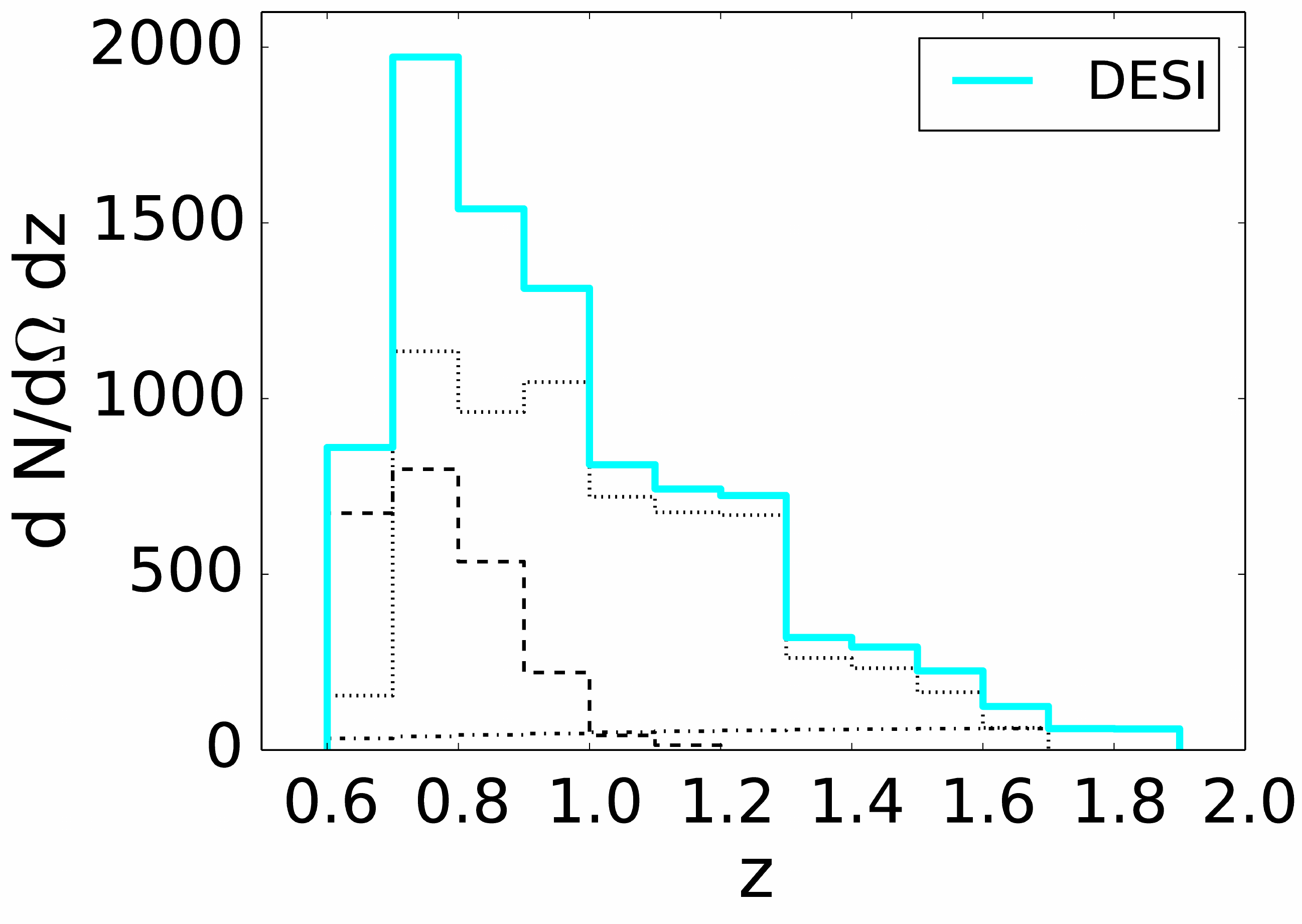}
\includegraphics[width=5.05 cm]{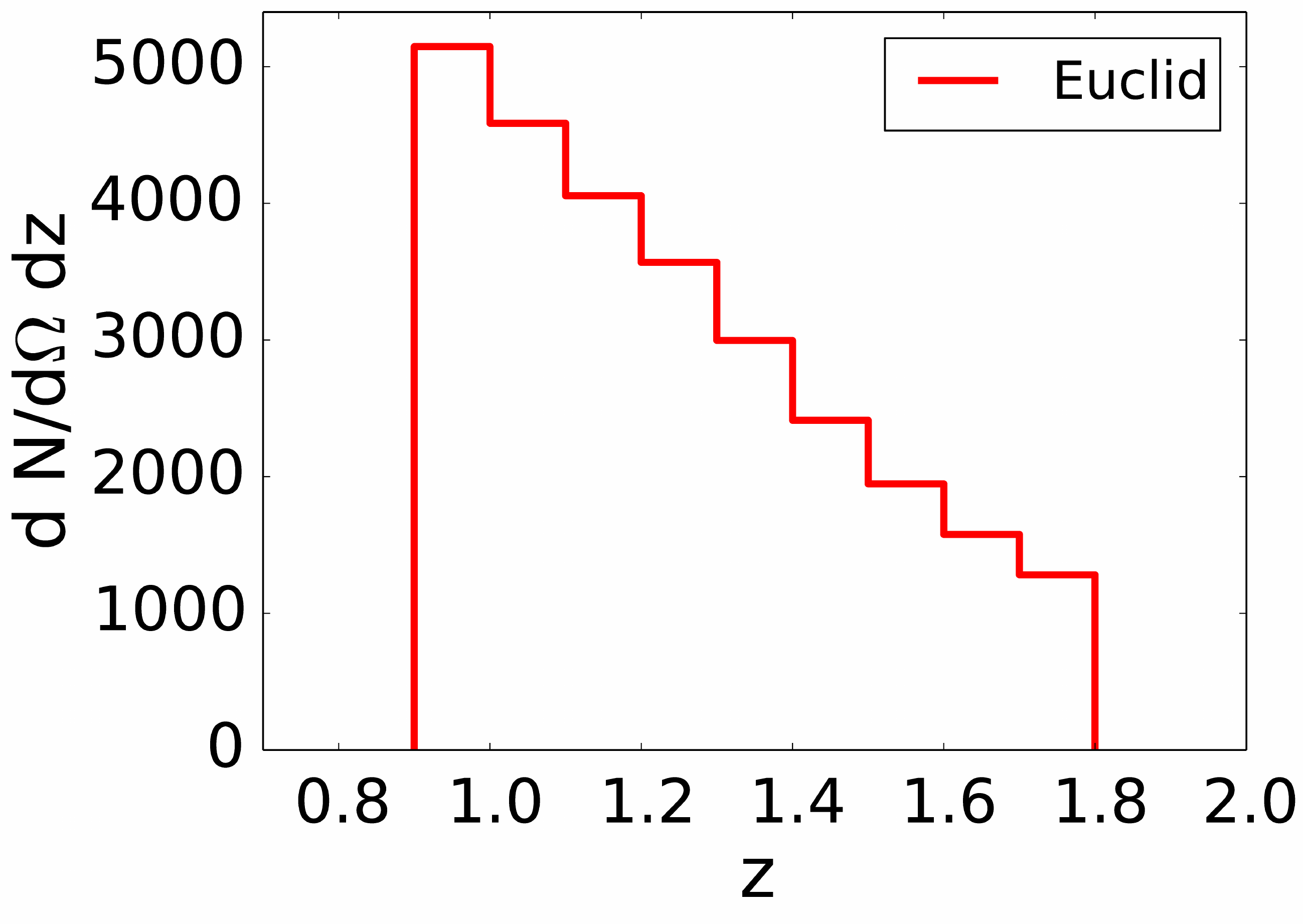}
\includegraphics[width=5.05 cm]{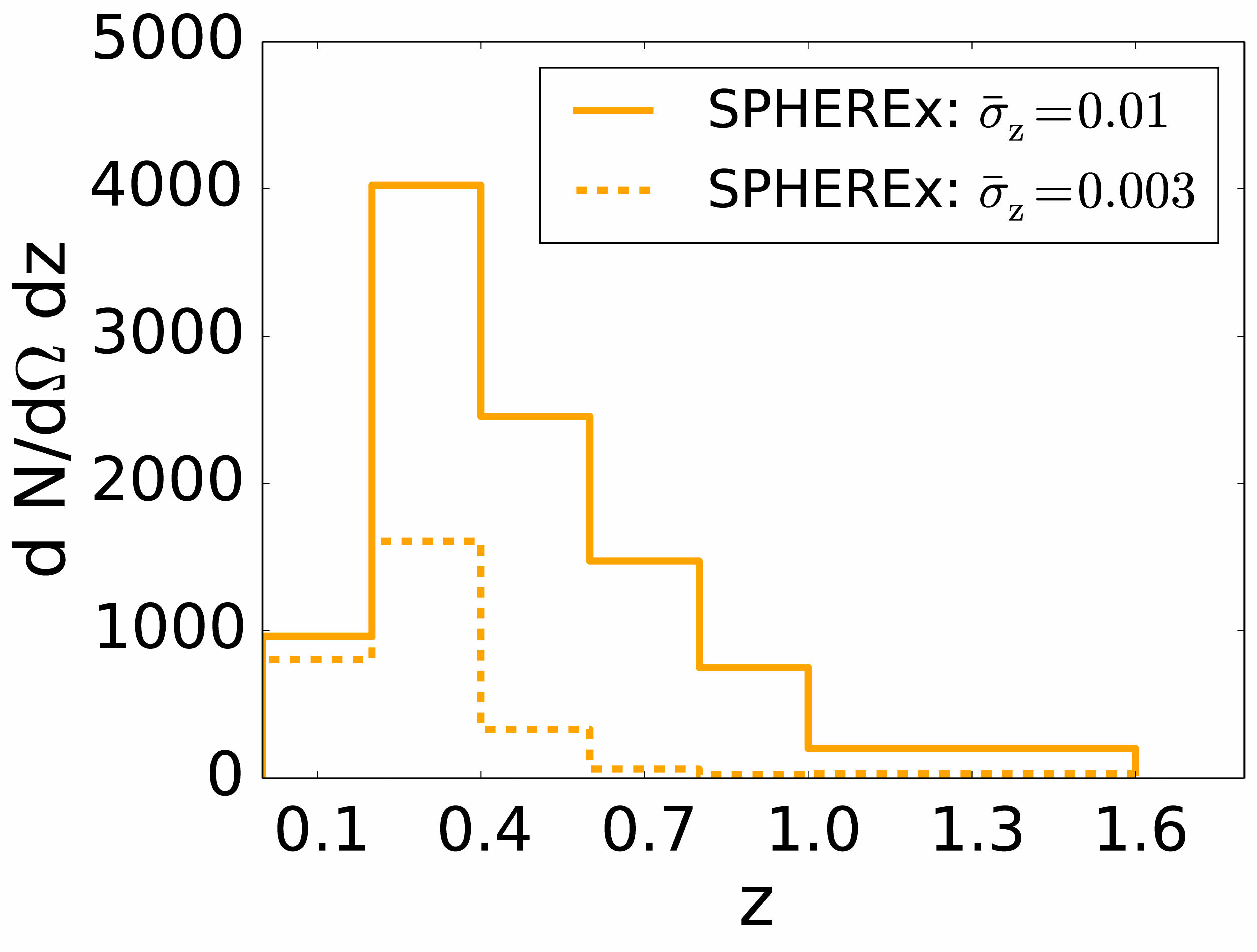}
\caption{Number density of objects, per unit of redshift bin $\Delta z$ and per square degree, used 
for the analysis. In the left panel we show the different tracers for DESI: LRGs (dashed), ELGs 
(dotted), QSOs (dot-dashed) and the total population (cyan) with a redshift bin $\Delta z = 0.1$. 
In the central panel we show the ELGs population expected with Euclid with a redshift bin 
$\Delta z = 0.1$. In the right panel we show two galaxy populations obtained for SPHEREx by 
considering different redshift uncertainty: the dashed (solid) line represent the observed 
objects considering a redshift uncertainty $\bar{\sigma}_z \sim 0.003$ (0.01) with a redshift bin 
of $\Delta z=0.2$ between 0 and 1 and of $\Delta z=0.6$ for higher redshifts}.\label{fig:samples}
\end{figure}

\begin{figure}[!ht]
\centering
\includegraphics[width=7 cm]{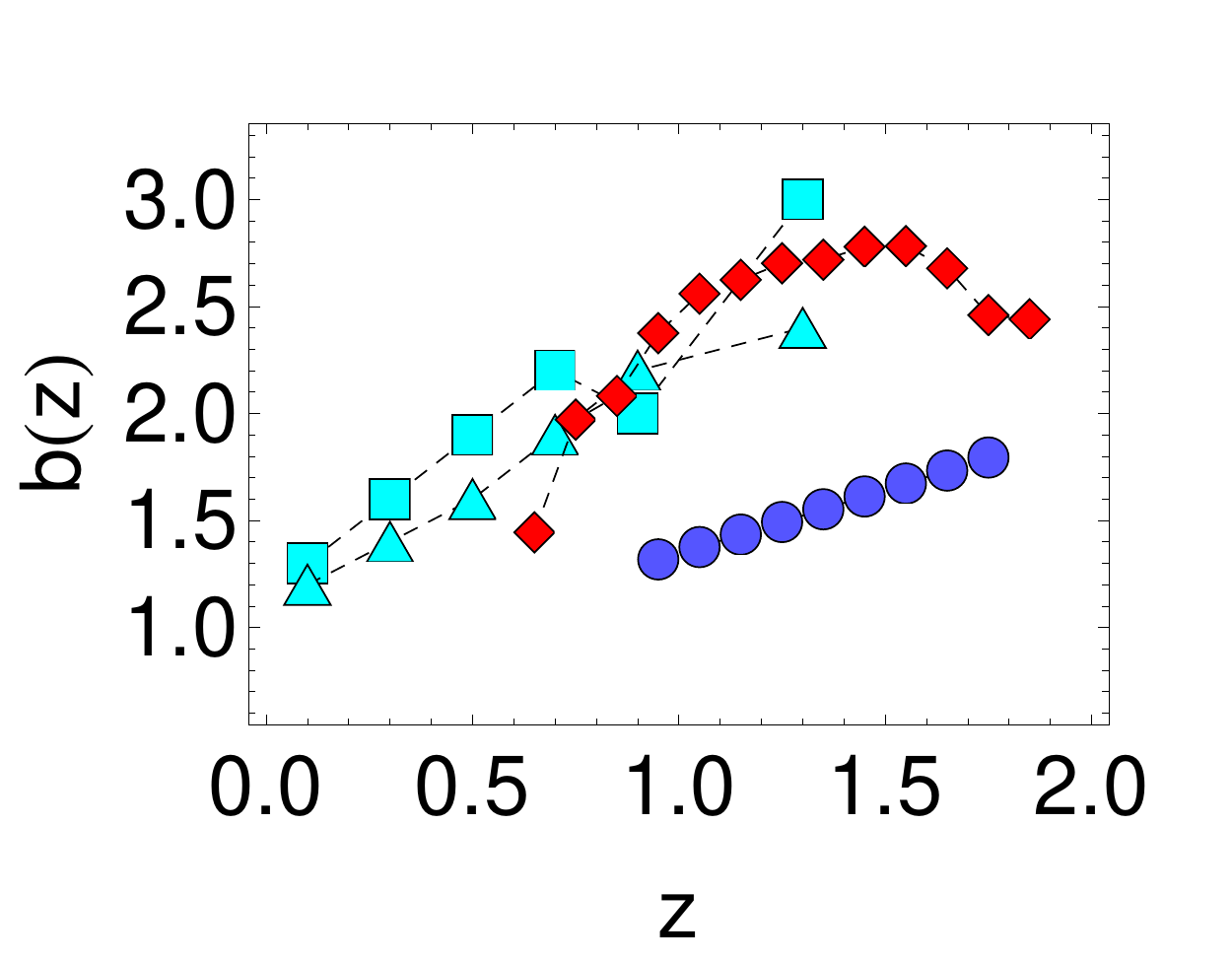}
\caption{We show the redshift evolution of the linear bias for each of the surveys considered: 
red diamonds for DESI, blue circles for Euclid, cyan triangles (squares) for SPHEREx1 
(SPHEREx2).}\label{fig:bias}
\end{figure}

\subsection{Dark Energy Spectroscopic Instrument (DESI)}

The DESI ground-based experiment \cite{Levi:2013gra} is expected to start observations in 2018 and 
to complete  in 4 years a 14000 deg$^2$ redshift survey of  galaxies and quasars. DESI will 
observe luminous red galaxies (LRGs) up to $z = 1.0$, it will target bright $[O II]$ emission line 
galaxies (ELGs) up to $z = 1.7$ and quasars (QSOs) up to $z < 2.1$. DESI will also  obtain a sample 
of bright galaxies at smaller redshifts ($0.05 < z < 0.4$) and one of higher-redshift 
($2.1 < z < 3.5$) quasars looking for the Lyman-$\alpha$ forest absorption features in their spectra. 
In our analysis we use for DESI the specifications from ref.~\cite{DESI:2015} (see in particular 
their table~2.3 and table~3.1).
We consider a combined galaxy clustering information for different tracers observed 
by DESI. In more detail, we use a simplified picture in which we assume that the different 
populations of LRGs, ELGs and QSOs are contributing to an effective unique population, covering 
thirteen redshift bins between $z$ = 0.6 and 1.9 with width of $\Delta z=0.1$, and having an effective 
bias given by \cite{Alonso:2015sfa}:
\be
b_\textup{eff}(z) = \frac{\bar{n}_{\rm LRG}(z)b_{\rm LRG}(z)
+\bar{n}_{\rm ELG}(z)b_{\rm ELG}(z)+\bar{n}_{\rm QSO}(z)b_{\rm QSO}(z)}
{\bar{n}_{\rm LRG}(z)+\bar{n}_{\rm ELG}(z)+\bar{n}_{\rm QSO}(z)}\,,
\ee
where we assume:
\begin{align}
&b_{\rm LRG}(z) = 1.7\ D(0)/D(z) \\
&b_{\rm ELG}(z) = 0.84\ D(0)/D(z) \label{eqn:biasELG} \\
&b_{\rm QSO}(z) = 1.34\ D(0)/D(z)\,.
\end{align}
 
This description is a good approximation of the exact multi-tracers approch in the limit of 
independent tracers \cite{Alonso:2015sfa}. For this purpose, we have also reduced the number of 
objects in the total sample as in \cite{Alonso:2015sfa} to include the effects of the target 
selection done to have a good redshift definition and to avoid confusion between different tracers 
and with other astrophysical objects (see the left panel of figure~\ref{fig:samples}). The 
resulting effective bias is shown in figure~\ref{fig:bias}.
As error for the DESI spectroscopic redshift we use $\bar{\sigma}_{\rm z} \sim 0.001$ \cite{DESI:2015}. 
As a reference, for DESI we obtain $k_{\rm min}$ ranging between $(3.59-4.71) \times 10^{-3}$ h 
Mpc$^{-1}$ for the different redshift bins here considered.

\subsection{Euclid}

The European Space Agency (ESA) Cosmic Vision mission Euclid \cite{Laureijs:2011gra} is scheduled to 
be launched in 2020, with the goal of characterising the dark sector of our Universe. This will be 
done mostly measuring the cosmic shear in a photometric surveys of billions of galaxies and the 
galaxy clustering in a spectroscopic survey of tens of millions of H$\alpha$ emitting galaxies. In 
this paper we will focus on the wide spectroscopic survey, which will cover an area of 15000 deg$^2$.

According to the updated predictions obtained by \cite{Pozzetti:2016cch}, the Euclid wide 
single-grism survey will reach a flux limit 
$F_{\textup{H}\alpha} > 2 \times 10^{-16}\ \text{erg}\ \text{cm}^{-2}\ \text{s}^{-1}$ 
and will cover a redshift range $0.9 < z < 1.8$. We consider nine redshift bins in this redshift 
range with the same width of $\Delta z=0.1$.
With these specifications and assuming a completeness 
of 70\%, the expected density number of H$\alpha$ emitters is about 4000 objects/deg$^2$, the 
redshift distribution of which (taken from table 3 of ref.~\cite{Pozzetti:2016cch}) is shown in the 
central panel of figure~\ref{fig:samples}. 
We can safely assume that the galaxy sample is composed by a single tracer, ELGs, and then assume that 
the bias follows eq.~\eqref{eqn:biasELG}.  Finally we adopt as redshift accuracy 
$\bar{\sigma}_{\rm z} \sim 0.001$ \cite{Amendola:2012ys}. As a reference, we obtain $k_{\rm min}$ in 
a range $(3.59-4.01) \times 10^{-3}$ h Mpc$^{-1}$ in the different redshift bins.

\subsection{Spectro-Photometer for the History of the Universe, Epoch of Reionization, and Ices Explorer (SPHEREx)}

SPHEREx \cite{Bock:2016,Dore:2014cca} is a NASA proposed small explorer satellite having the goal 
of providing the first near-infrared spectro-photometric image of the complete sky, thanks to its 
coverage of 40000 deg$^2$ in the wavelength range $0.75<\lambda\ \mu\text{m}^{-1}<4.8$.

SPHEREx will collect spectra of galaxies at $z<1$, covering the redshift range for clustering 
studies that are not covered by the Euclid spectroscopic survey. Moreover it will observe 
high-redshift quasars in its deep fields.
In our analysis we will consider only the galaxy sample, assuming that the fraction of sky usable 
for clustering studies is 75\% of the whole sky, in strict analogy to what is done in CMB analyses.

We consider two different configurations for SPHEREx \footnote{We wish to thank Olivier Dor\'e and 
Roland de Putter for making available the SPHEREx specifications to us.}, 
with $\bar{\sigma}_{\rm z} \sim 0.003$ (hereafter SPHEREx1) and $\bar{\sigma}_{\rm z} \sim 0.01$ 
(hereafter SPHEREx2).

For the two different configurations we consider five redshift bins, between $z$ = 0.0 and 
1.0, and one redshift bin, between $z$ = 1.0 and 1.6, with a width 
of $\Delta z=0.2$ between 0 and 1 and of $\Delta z=0.6$ for higher redshifts.
The adopted bias is shown in figure~\ref{fig:bias}.
As a reference, for SPHEREx we obtain $k_{\rm min}$ in the range 
$(1.60-7.51) \times 10^{-3}$ h Mpc$^{-1}$.

\section{Results and Discussions}
\label{sec:results}

We now discuss the uncertainties in the cosmological parameters obtained as result of our combined 
CMB and LSS Fisher approach.

For the $\Lambda$CDM model the uncertainties in the cosmological parameters  are reported in 
table~\ref{tab:resultsCMB}. Our results for the uncertainties from CMB and LSS are broadly 
consistent with the previous ones in the literature \cite{Huangetal,Basse:2014qqa}. 
We need however to bear in mind that different assumptions for CMB specifications 
were considered in ref.~\cite{Huangetal} and ref.~\cite{Basse:2014qqa}.

\begin{center}
\begin{table*}[ht]
\centering{\scriptsize
\caption{Forecasts for the marginalized 68\% uncertainties for the cosmological parameters in 
the $\Lambda$CDM model with our Fisher approach.
The two results in the parentheses include the constraints obtained in combination with 
the CMB Fisher matrix for the two configurations (CMB-1 and CMB-2, respectively). 
We do not list the forecasted uncertainty for $\tau$ since it does not benefit from the 
inclusion of LSS.}
\vspace{3mm}
\begin{tabular}{|l|cccc|}
\hline
 & DESI & Euclid & SPHEREx1 & SPHEREx2 \\
\hline
$10^3\ \sigma\left(\Omega_{\rm c}\right)$ 		      & 11.6 (2.6/2.5)   & 9.6  (2.1/2.0)   & 13.1 (2.9/2.7)   & 7.1  (1.8/1.6) \\
$10^3\ \sigma\left(\Omega_{\rm b}\right)$ 		      & 4.1  (0.28/0.26) & 3.0  (0.25/0.23) & 4.6  (0.30/0.29) & 2.5  (0.23/0.21)\\
$\sigma\left(H_0\right)$        		              & 4.0  (0.21/0.20) & 3.0  (0.17/0.16) & 4.4  (0.23/0.22) & 2.6  (0.15/0.14)\\
$10^2\ \sigma\left(n_{\rm s}\right)$       		      & 6.7  (0.26/0.24) & 5.3  (0.25/0.22) & 7.5  (0.26/0.23) & 4.2  (0.24/0.22)\\
$10^2\ \sigma\left(\ln\left(10^{10}A_{\rm s}\right)\right)$   & 35.5 (0.80/0.71) & 32.9 (0.74/0.67) & 37.9 (0.83/0.74) & 21.2 (0.74/0.67)\\
\hline

\end{tabular}}
\label{tab:resultsCMB}
\end{table*}
\end{center}
\begin{figure}[!ht]
\centering
\includegraphics[width=7.6cm]{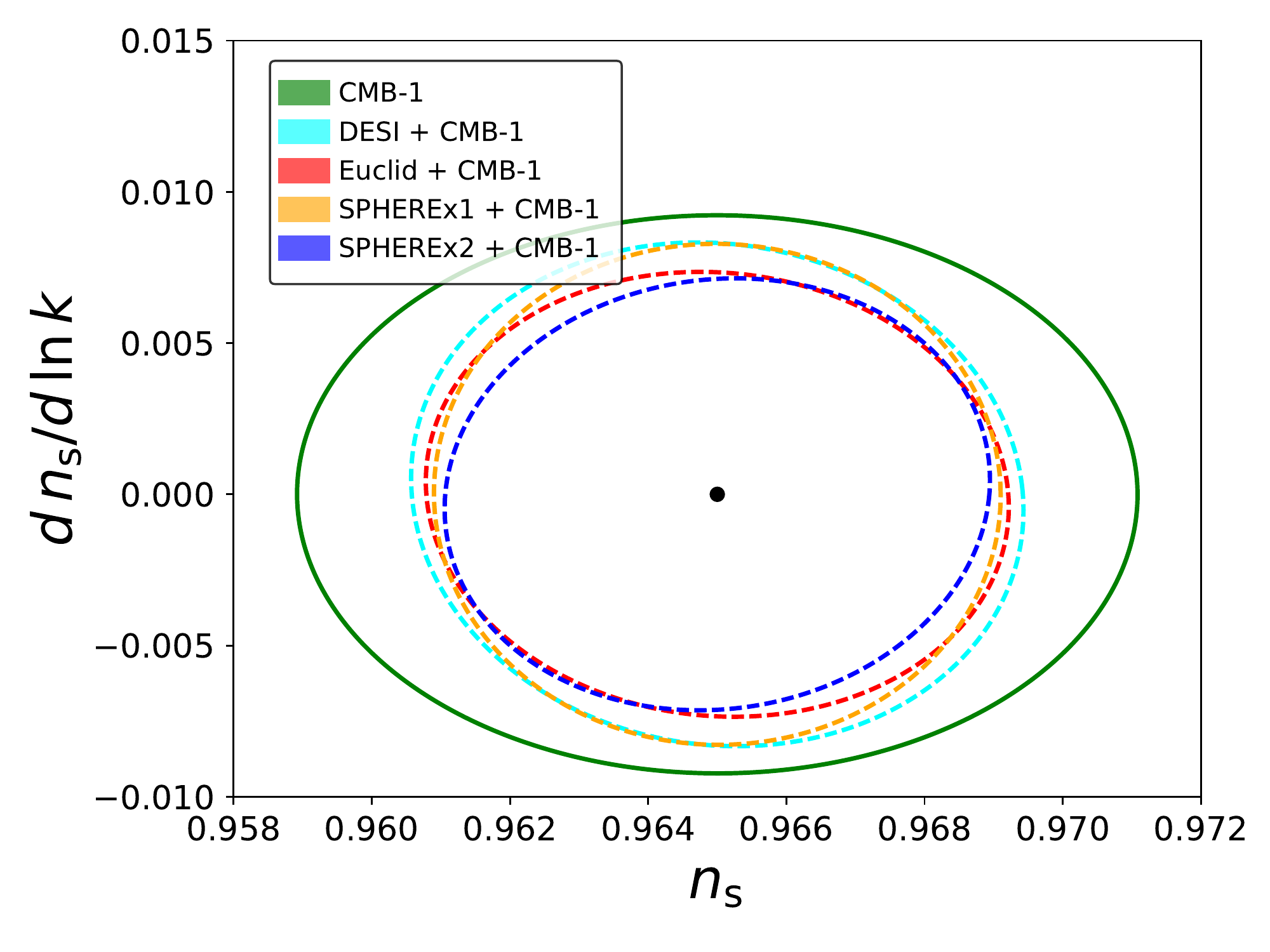} \includegraphics[width=7.6cm]{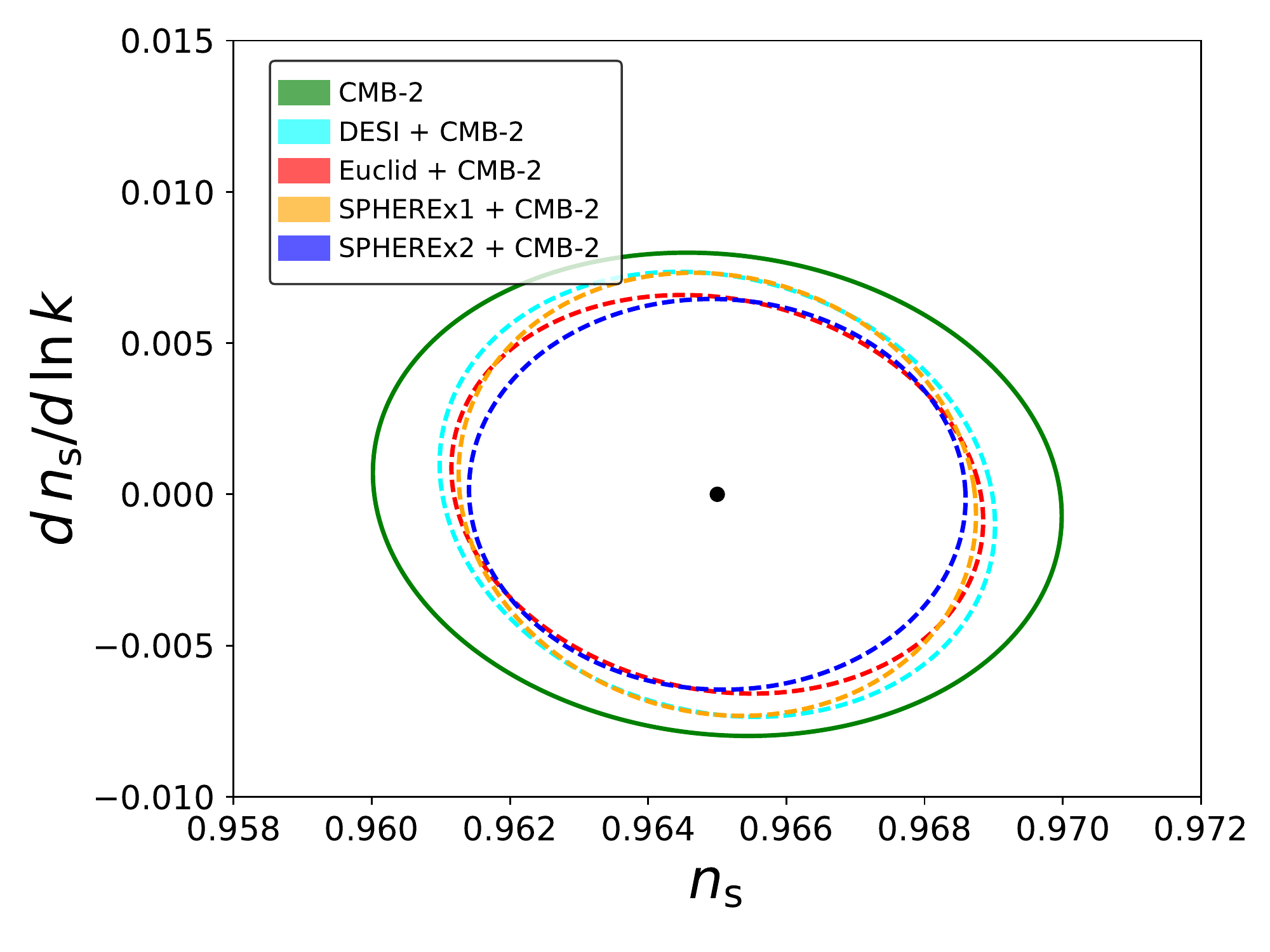}
\caption{CMB and LSS combined constraints on 
$(n_{\rm s}$, ${\rm d}  n_{\rm s} /{\rm d} \ln k)$ at 68\% CL. \textcolor{red}{Different lines refer to
CMB only (solid green), DESI (dashed cyan), Euclid (dashed red), SPHEREx (dashed orange) 
and SPHEREx (dashed oblue).
The contours for the LSS surveys are in combination with CMB-1 (CMB-2) in the left (right) panel.}}\label{fig:scalarunning}
\end{figure}

We have also analyzed the case in which the dependence in the wavelength of the spectral index 
is allowed to vary, by fixing the fiducial model to 
$(n_{\rm s}, {\rm d} n_{\rm s} /{\rm d} \ln k) = (0.9655, 0.0)$. We obtain the following 
uncertainties $(\sigma (n_{\rm s}), \sigma ({\rm d} n_{\rm s} /{\rm d} \ln k) )$: 
$(0.0029,0.0053)$ for DESI, 
$(0.0027,0.0047)$ for Euclid,
$(0.0027,0.0039)$/$(0.0026,0.0036)$ for SPHEREx1/SPHEREx2,
when the CMB-1 Fisher information for the more conservative configuration is added.
When combining the Fisher information for the second CMB configuration with the LSS one, the errors 
are slightly decreased as can be seen in figure~\ref{fig:scalarunning}. 
Being ${\rm d} n_{\rm s} /{\rm d} \ln k = - 0.003 \pm 0.007$ the current $Planck$ measurement on the 
running \cite{PlanckXX2015,Ade:2015xua}, the parameter space with ${\rm d} n_{\rm s} /{\rm d} \ln k$ 
exceeding the standard slow-roll predictions $\approx (n_{\rm s}-1)^2$ will be further probed by 
future galaxy surveys.

Geometrical distortions to the galaxy power spectrum due to the changes in $H(z)$ and 
$D_{\rm A}(z)$ will cause both a horizontal and vertical shift in the observed power spectrum 
and introduce new degeneracies in the measured power spectrum \cite{Shoji:2008xn}. 
The AP effect instead has a main impact on the late-time parameters \cite{Bailoni:2016ezz}.

Overall, the impact of the geometrical distortions and of the AP term included in the analysis, 
see Eqs.~\eqref{eqn:obsPdk}-\eqref{eqn:kref}-\eqref{eqn:muref}, mainly affect the 
uncertainties of the standard cosmological parameters of 
the $\Lambda$CDM model and to a smaller extent the running of the spectral index.
They have a small impact on the uncertainties of the extra parameters of the models with 
features in the PPS, in particular after having marginalized over the several nuisance parameters.

We now discuss our results for the four inflationary models with features considered. The results 
are summarized in table~\ref{tab:sigma}.
\begin{center}
\begin{table*}
\centering\scriptsize{
\caption{Forecasts for the marginalized 68\% uncertainties for the features parameters for any survey considered in combination with CMB-1 (CMB-2 in parenthesis).}
\vspace{3mm}
\makebox[0pt]{
\begin{tabular}{|c|l c|c|c|c|c|}
\hline
Model & Parameter            & & DESI            & Euclid          & SPHEREx1        & SPHEREx2\\
      & (Best-fit)           & & + CMB-1 (CMB-2) & + CMB-1 (CMB-2) & + CMB-1 (CMB-2) & + CMB-1 (CMB-2)\\
\hline
\multirow{2}{*}{MI} & $\lambda_{\rm c} $ & (0.5)                                & $0.22\ (0.21)$ & $0.22\ (0.21)$ & $0.22\ (0.22)$ & $0.21\ (0.20)$\\
\cline{2-7}
                    & $\log_{10} \left(k_{\rm c}\ {\rm Mpc} \right) $ & (-3.47) & $0.40\ (0.39)$ & $0.39\ (0.39)$ & $0.41\ (0.40)$ & $0.38\ (0.38)$\\
\hline
\multirow{2}{*}{MII} & $\Delta$ & (0.089)                                       & $0.034\ (0.033)$ & $0.031\ (0.040)$ & $0.034\ (0.033)$ & $0.027\ (0.026)$\\
\cline{2-7}
                     & $\log_{10} \left(k_{\rm s}\ {\rm Mpc} \right)$ & (-3.05) & $0.077\ (0.071)$  & $0.071\ (0.066)$ & $0.079\ (0.072)$  & $0.061\ (0.057)$\\
\hline
\multirow{3}{*}{MIII} & $\mathcal{A}_{\rm st}$ & (0.374)                          & $0.24\ (0.22)$   & $0.24\ (0.22)$   & $0.24\ (0.22)$   & $0.20\ (0.19)$\\ 
\cline{2-7}
                      & $\log_{10} \left(k_{\rm st}\ {\rm Mpc} \right)$ & (-3.10) & $0.038\ (0.026)$ & $0.033\ (0.024)$ & $0.031\ (0.027)$ & $0.024\ (0.022)$\\
\cline{2-7}
                      & $\ln x_{\rm st}$ & (0.342)                                & $0.34\ (0.30)$   & $0.32\ (0.29)$   & $0.33\ (0.30)$   & $0.28\ (0.26)$\\
\hline
\multirow{3}{*}{MIV} & $\mathcal{A}_{\rm log}$ & (0.0278)                  & $0.0035\ (0.0032)$ & $0.0030\ (0.0028)$ & $0.0038\ (0.0035)$ & $0.0025\ (0.0024)$\\
\cline{2-7}
                     & $\log_{10} \left(\omega_{\rm log} \right)$ & (1.51) & $0.0087\ (0.0079)$ & $0.0077\ (0.0071)$ & $0.0094\ (0.0084)$ & $0.0062\ (0.0059)$\\
\cline{2-7}
                     & $\phi_{\rm log}/(2\pi)$ & (0.634)                   & $0.020\ (0.018)$   & $0.017\ (0.016)$   & $0.021\ (0.019)$   & $0.014\ (0.013)$\\
\hline
\end{tabular}}}
\label{tab:sigma}
\end{table*}
\end{center}

\begin{figure}[!ht]
\centering
\includegraphics[width=7.6cm]{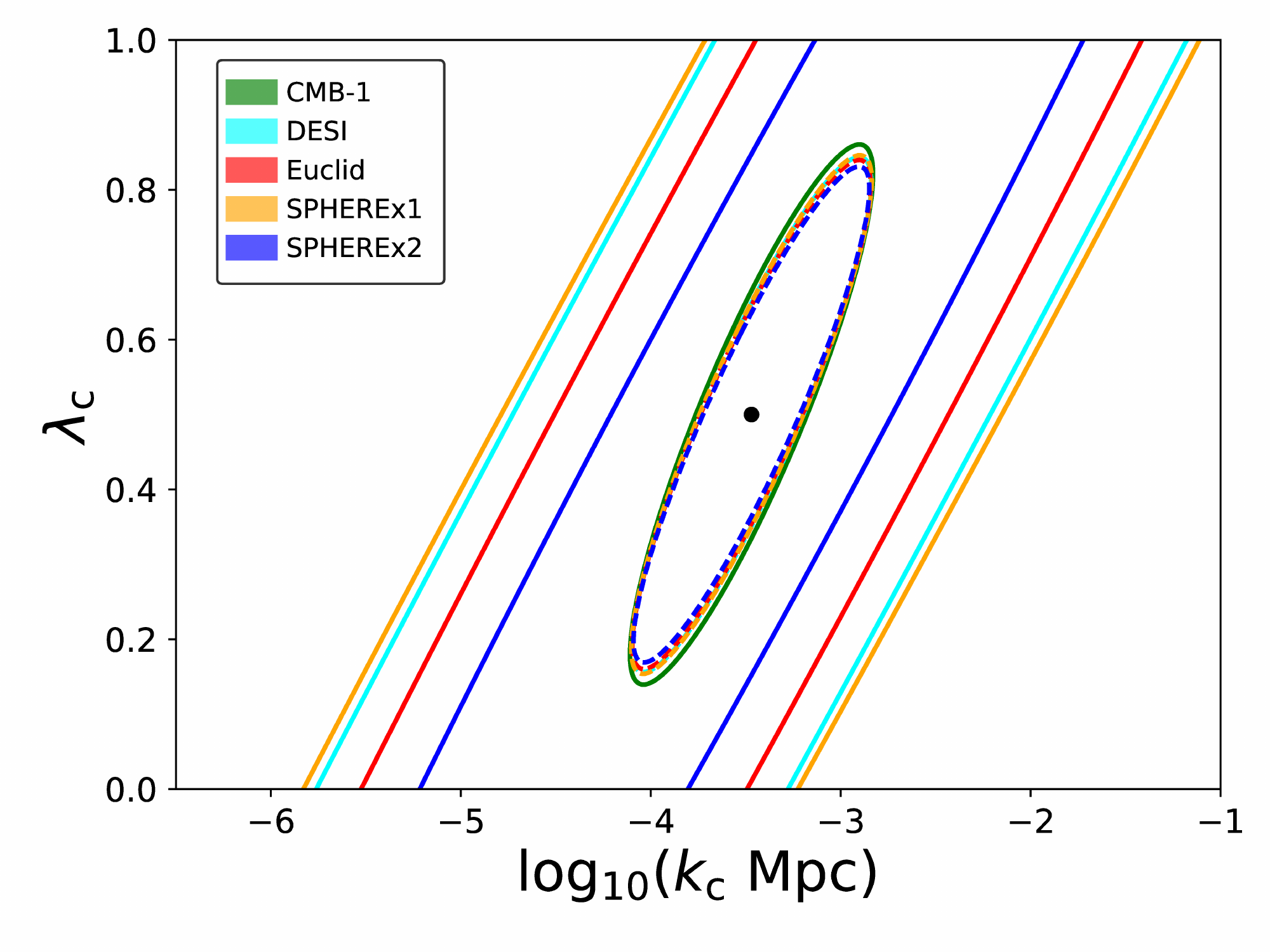} \includegraphics[width=7.6cm]{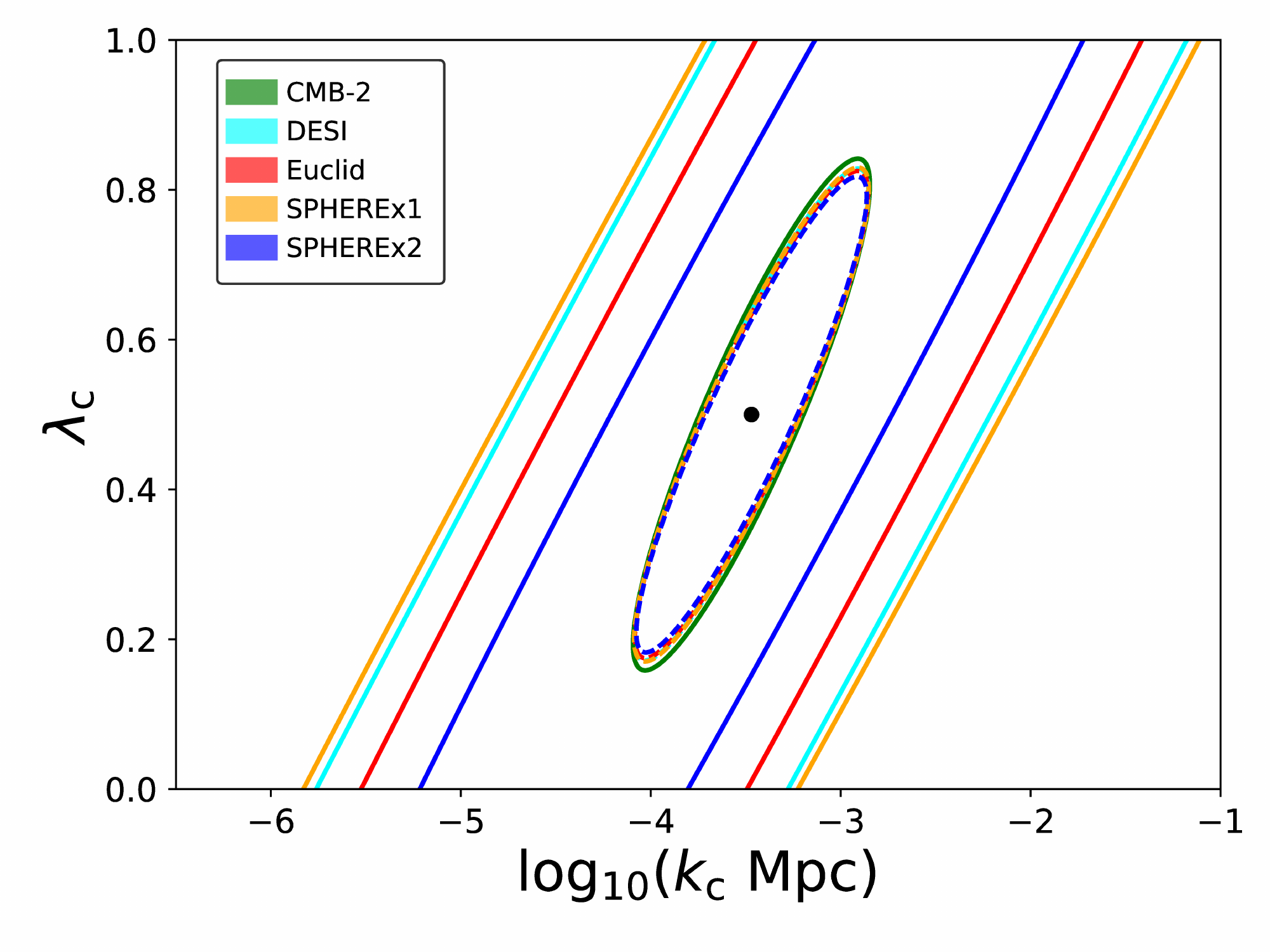}
\caption{Marginalized 2D 68\% CL contours for the parameters 
$(\log_{10}(k_c\ $Mpc$)$, $\lambda_c)$ of MI for CMB only (solid green), DESI 
(solid cyan), Euclid (solid red), SPHEREx1 (solid orange), and SPHEREx2 (solid blue). 
The dashed contours represent the 2D 68\% CL CMB and LSS combined results. The configuration 
CMB-1 (CMB-2) is considered in the left (right) panel.\label{fig:cutoff}}
\end{figure}

The effective very large scale of MI obtained as a best-fit for $Planck$ 2015 \cite{PlanckXX2015} is 
a challenge for the future galaxy surveys here considered (see figure~\ref{fig:cutoff}).
Such a modification on large scales seems a better target for high-sensitivity CMB polarization 
experiments covering a large fraction of the sky, such as $Planck$, 
AdvACTpol~\cite{Henderson:2015nzj}, CLASS~\cite{Essinger-Hileman:2014pja}, LSPE~\cite{:2012goa}, 
which will provide an improved measurement of the E-mode polarization on large scales.
Note that the same type of suppression of this model has been previously studied in 
\cite{Gibelyouetal}: however, the fiducial cosmological model in ref.~\cite{Gibelyouetal} has 
been taken with $k_{\rm c} = 9.5\times 10^{-4}$ Mpc$^{-1}$, i.e. a wavenumber which is almost 
three times larger than the one suggested by the latest $Planck$ data, and with a much steeper 
cut-off, i.e. $\lambda_{\rm c} = 3$. 
The parametrized suppression of PPS chosen in \cite{Gibelyouetal} would be therefore a much 
easier target for future galaxy surveys.

\begin{figure}[!ht]
\centering
\includegraphics[width=7.6cm]{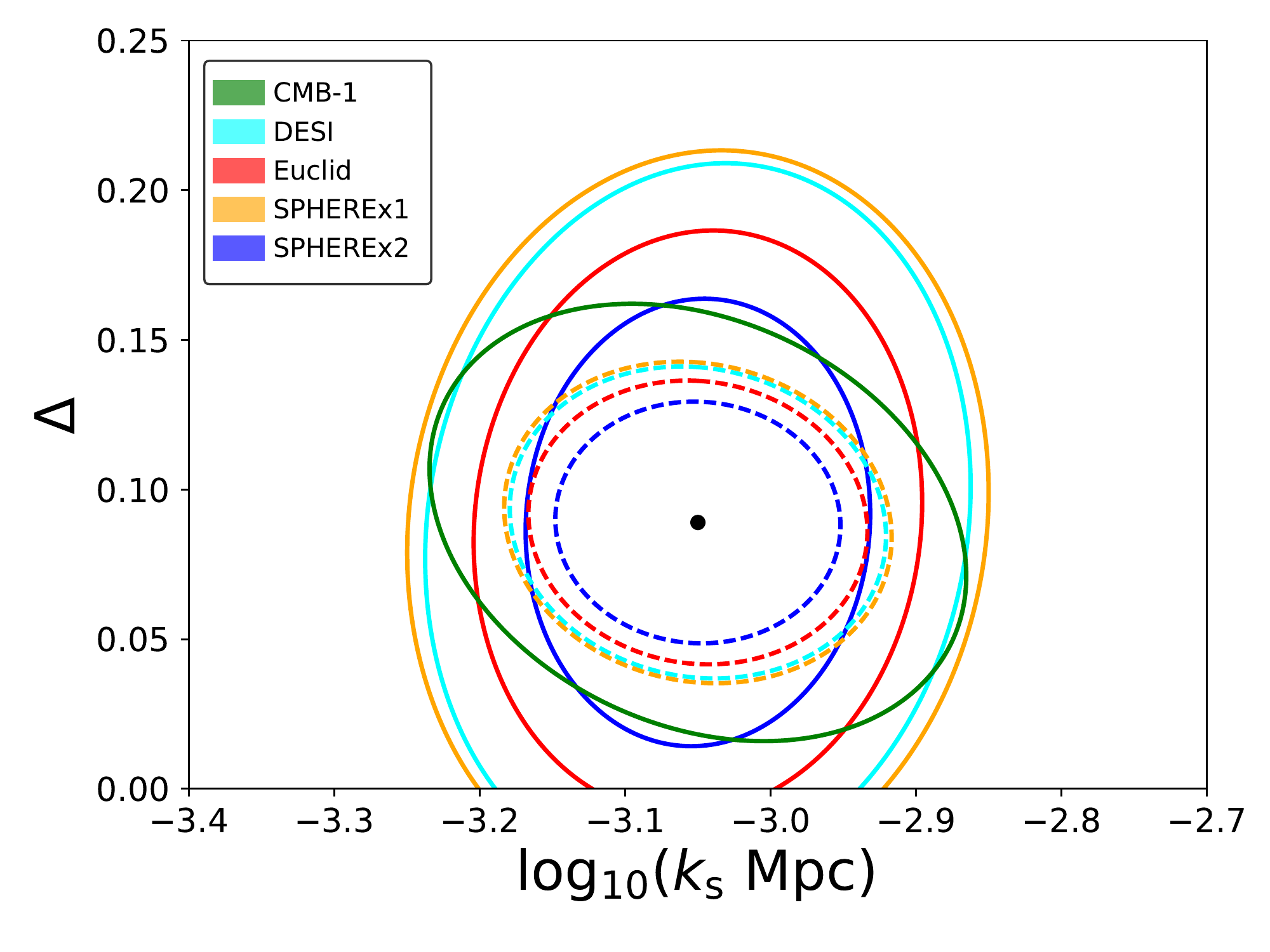} \includegraphics[width=7.6cm]{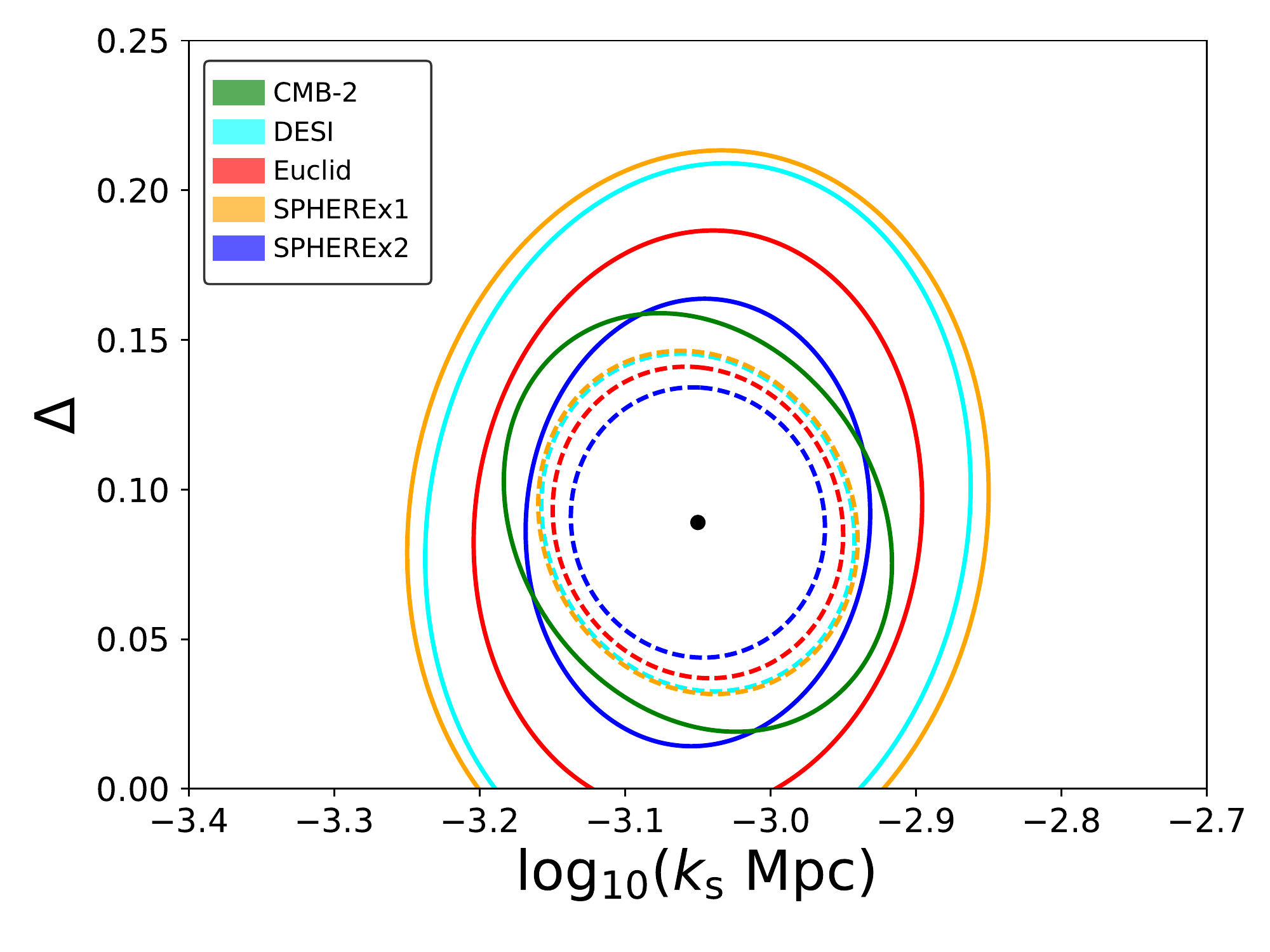}
\caption{Marginalized 2D 68\% CL contours for the parameters 
$(\log_{10}(k_s\ $Mpc$)$, $\Delta)$ of MII for CMB only (green), DESI (solid cyan), 
Euclid (solid red), SPHEREx1 (solid orange), and SPHEREx2 (solid blue). 
The dashed contours represent the 2D 68\% CL CMB and LSS combined results. The configuration 
CMB-1 (CMB-2) is considered in the left (right) panel.}\label{fig:staro}
\end{figure}

The model MII, with a discontinuity in the first derivative of the potential \cite{Starobinsky:1992ts}, 
has also two parameters as the first model, but the resulting power spectrum has super-imposed 
oscillations accompanying the change in the amplitude of the PPS. These oscillations are non-zero 
at scales smaller than the change in amplitude and can be therefore a target for future galaxy 
surveys. Whereas CMB is sensitive to the preferred scale of the model, the matter power spectrum 
from galaxy surveys is also more sensitive to the change in the amplitude of the power spectrum: for 
this model the complementarity of CMB and LSS is quite striking. As from figure~\ref{fig:staro}, 
the scale of the feature would be probed at higher statistical significance.
Also in this case, a previous study \cite{Huangetal} considered the capability of Euclid when 
combined with $Planck$ to discriminate this model for different values of the parameters. 
However, in \cite{Huangetal} a  comoving scale $k_{\rm s} = 6.8\times 10^{-2}$ Mpc$^{-1}$, i.e.  
8 times larger than the one suggested  by $Planck$ 2015 data and used here, was considered. Again, 
such a choice would enhance the possibility of detecting the feature either in CMB and LSS.

\begin{figure}[!ht]
\centering
\includegraphics[width=7.6cm]{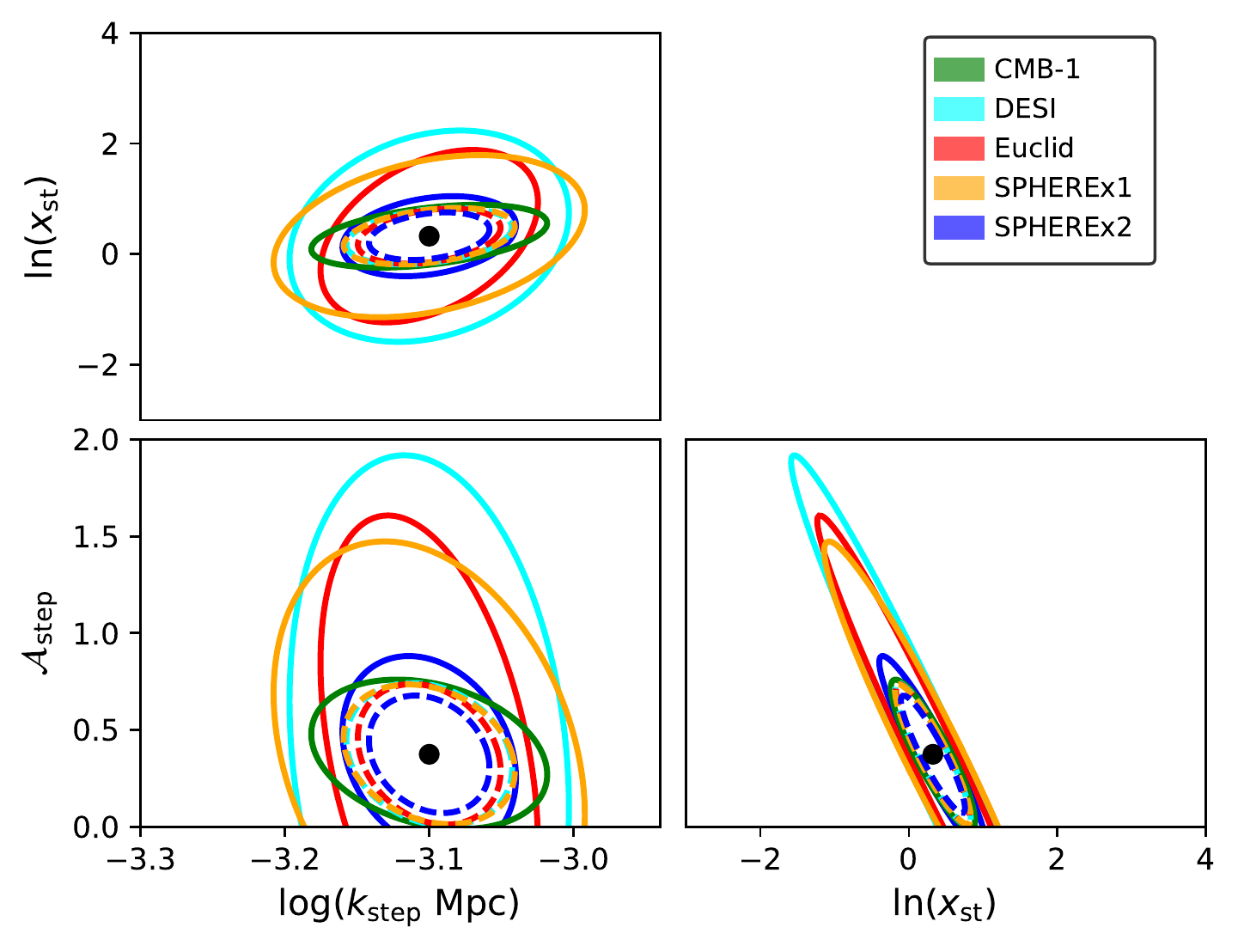} \includegraphics[width=7.6cm]{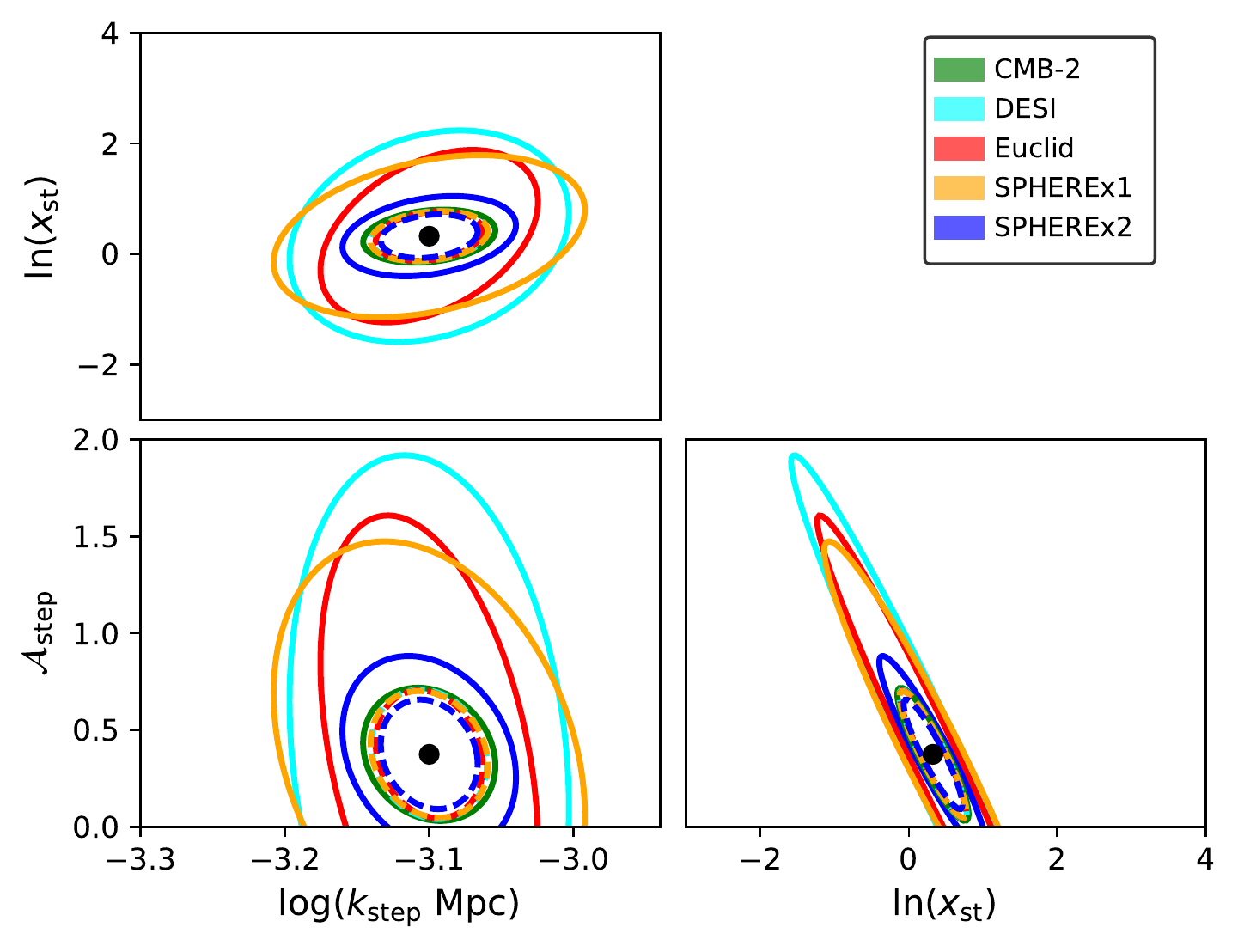}
\caption{Triangle plot with marginalized 2D 68\% CL contours for the parameters 
($\mathcal{A}_{\rm st}$, $\log_{10}(k_{\rm st}\ $Mpc$)$, $\ln(x_{\rm st})$) of MIII for CMB only 
(solid green), DESI (solid cyan), Euclid (solid red), SPHEREx1 (solid orange), 
and SPHEREx2 (solid blue). 
The dashed contours represent the 2D 68\% CL CMB and LSS combined results. 
The configuration CMB-1 (CMB-2) is considered in the left (right) panel.}\label{fig:step}
\end{figure}

The model with a step in the potential (MIII) benefits from the addition of LSS, as it can be seen from 
figure~\ref{fig:step}. In this case the power spectrum of galaxy surveys is sensitive to either 
the amplitude and the width of the ringing features in the primordial fluctuations; again, the scale 
of the feature would be probed at high statistical significance. 

For the fourth model considered, CMB and LSS can probe the amplitude of periodic oscillations at 
high statistical significance: we obtain $\mathcal{A}_{\rm log} = 0.0278 \pm 0.0030$ 
($\mathcal{A}_{\rm log} = 0.0278 \pm 0.0028$) at 68\% for CMB-1 (CMB-2) combined with Euclid.
This parameterization was also studied in \cite{Huangetal} but considering a different best-fit with 
a smaller amplitude and a frequency of $\omega_{\rm log} \sim 10$. 
Even if the constraint from CMB only in \cite{Huangetal} is tighter than the one we find, the 
improvement from CMB and Euclid in \cite{Huangetal} is in agreement with our finding.
We also checked that our fiducial frequency, $\omega_{\rm log} \sim 32$, does not disappear in 
$\ell$-space (keeping the frequency fixed) due the acoustic transfer function. By decreasing the 
amplitude of the periodic oscillations, the relative weight of the LSS increases with respect to CMB 
in the combined constraints; we have explicitly checked that half of the amplitude can still be 
detected at $3 \sigma$ by CMB-2 + Euclid (see also fig.~\ref{fig:wigg2}).

\begin{figure}[!ht]
\centering
\includegraphics[width=7.6cm]{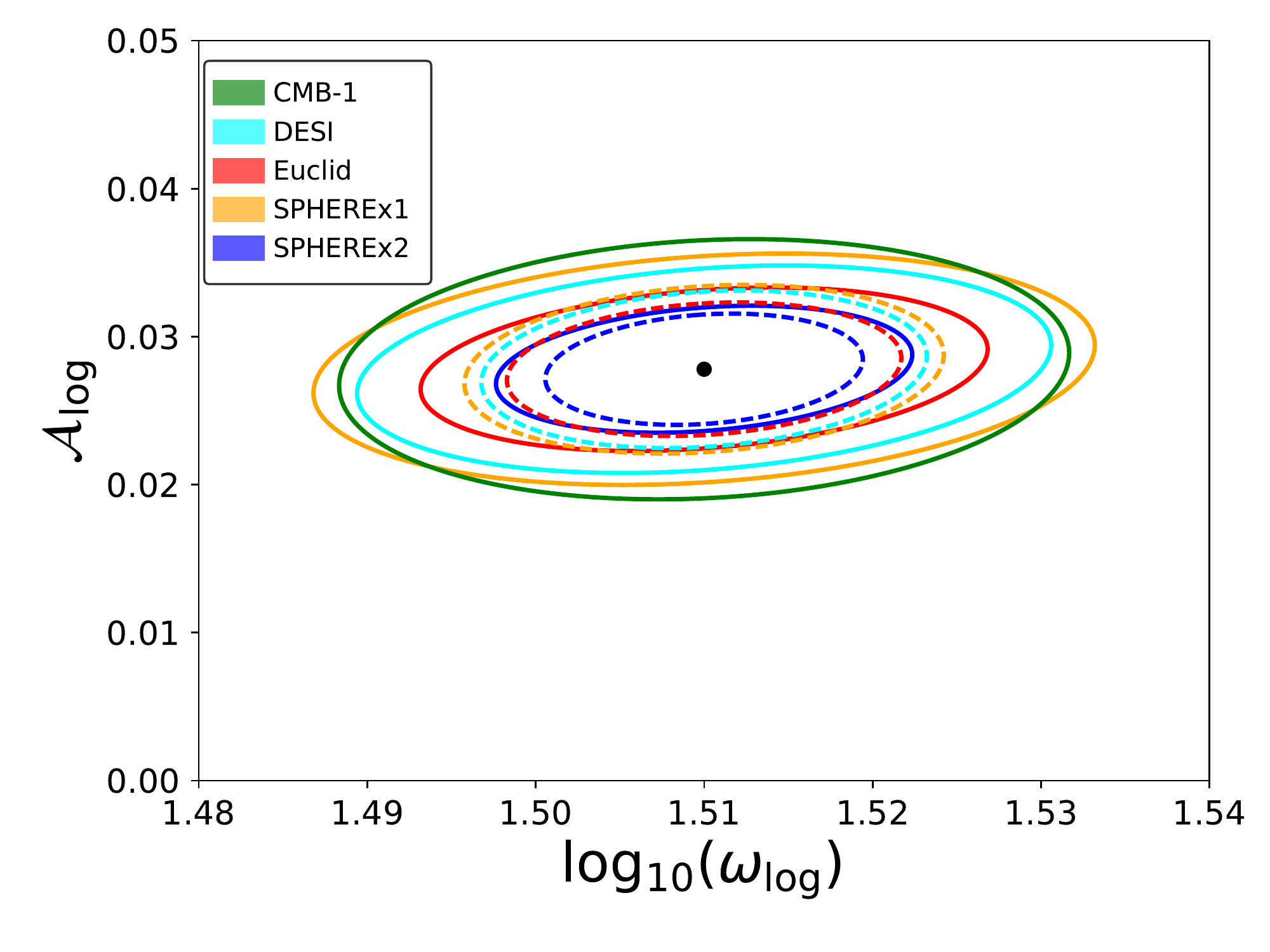} \includegraphics[width=7.6cm]{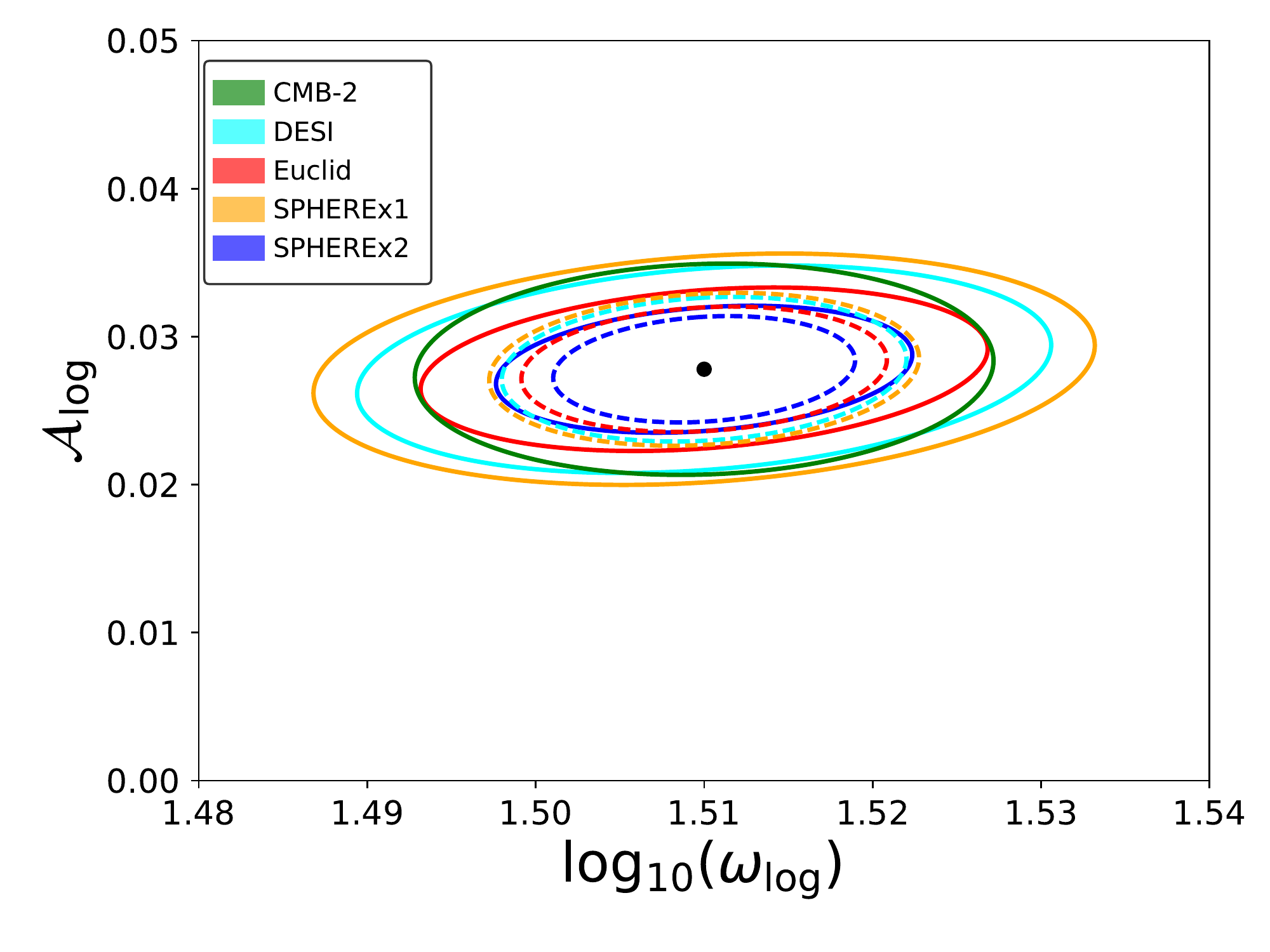}
\caption{Marginalized 2D 68\% CL contours for the parameters 
$(\log_{10}(\omega_{\rm log})$, $\mathcal{A}_{\rm log})$ of MIV for CMB only (green), 
DESI (solid cyan), Euclid (solid red), SPHEREx1 (solid orange), and SPHEREx2 (solid blue). 
The dashed contours represent the 2D 68\% CL CMB and LSS combined results. The configuration 
CMB-1 (CMB-2) is considered in the left (right) panel.}\label{fig:wigg}
\end{figure}

We stress that we have considered discrete bins linearly spaced for $P(k)$ with the minimum width 
(for every redshift slice) such as a diagonal covariance matrix is a good approximation 
\cite{Abramo:2011ph}. We believe that this setting is more conservative than considering a continuous 
$P(k)$ in the galaxy likelihood evaluation, given three of the considered fiducial models have 
super-imposed oscillations. If we were considering a continuous $P(k)$, there would be no considerable 
changes for MI, but we would obtain tighter constraints for the other three models.

\begin{figure}[!ht]
\centering
\includegraphics[width=7.6cm]{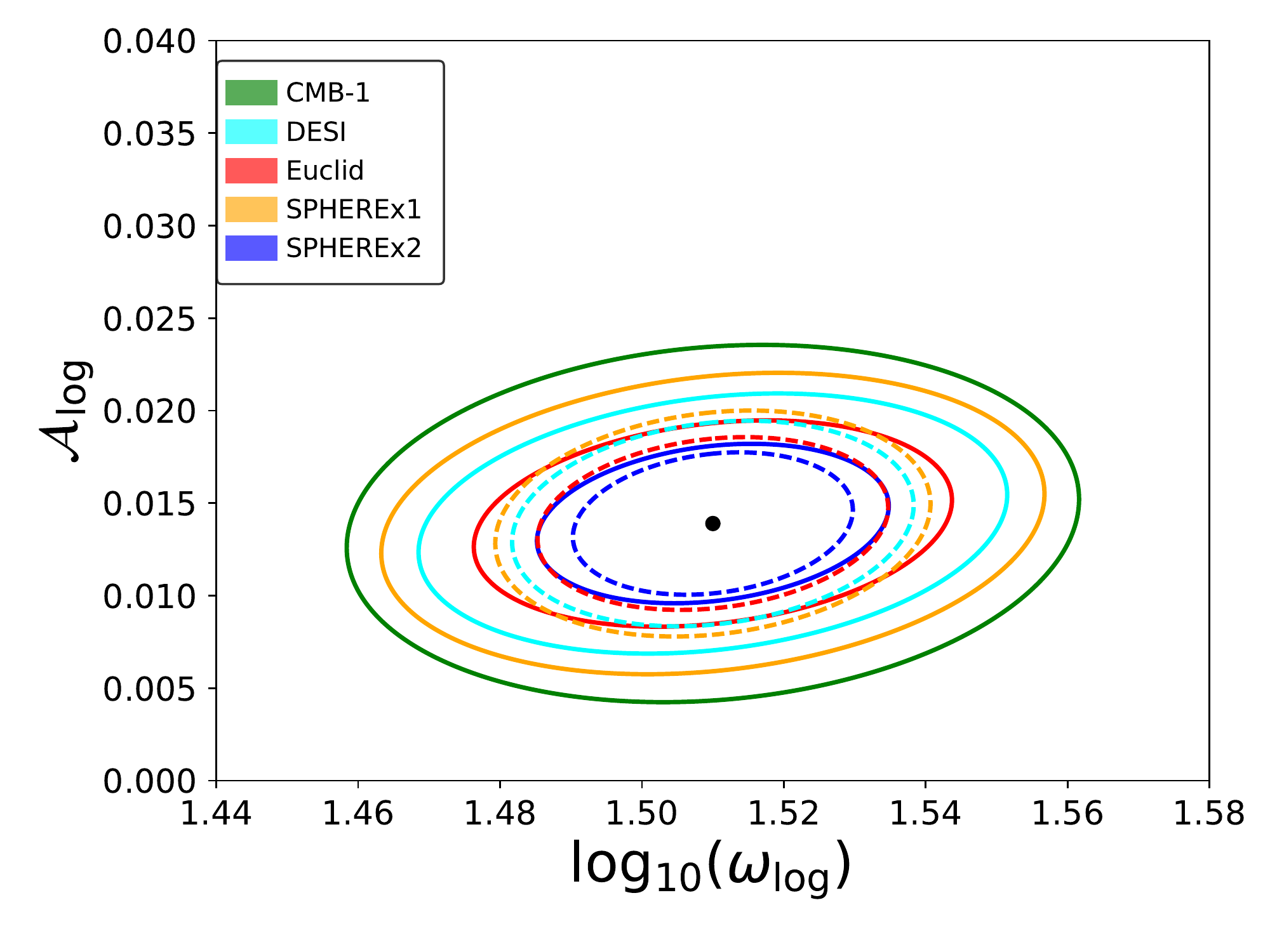} \includegraphics[width=7.6cm]{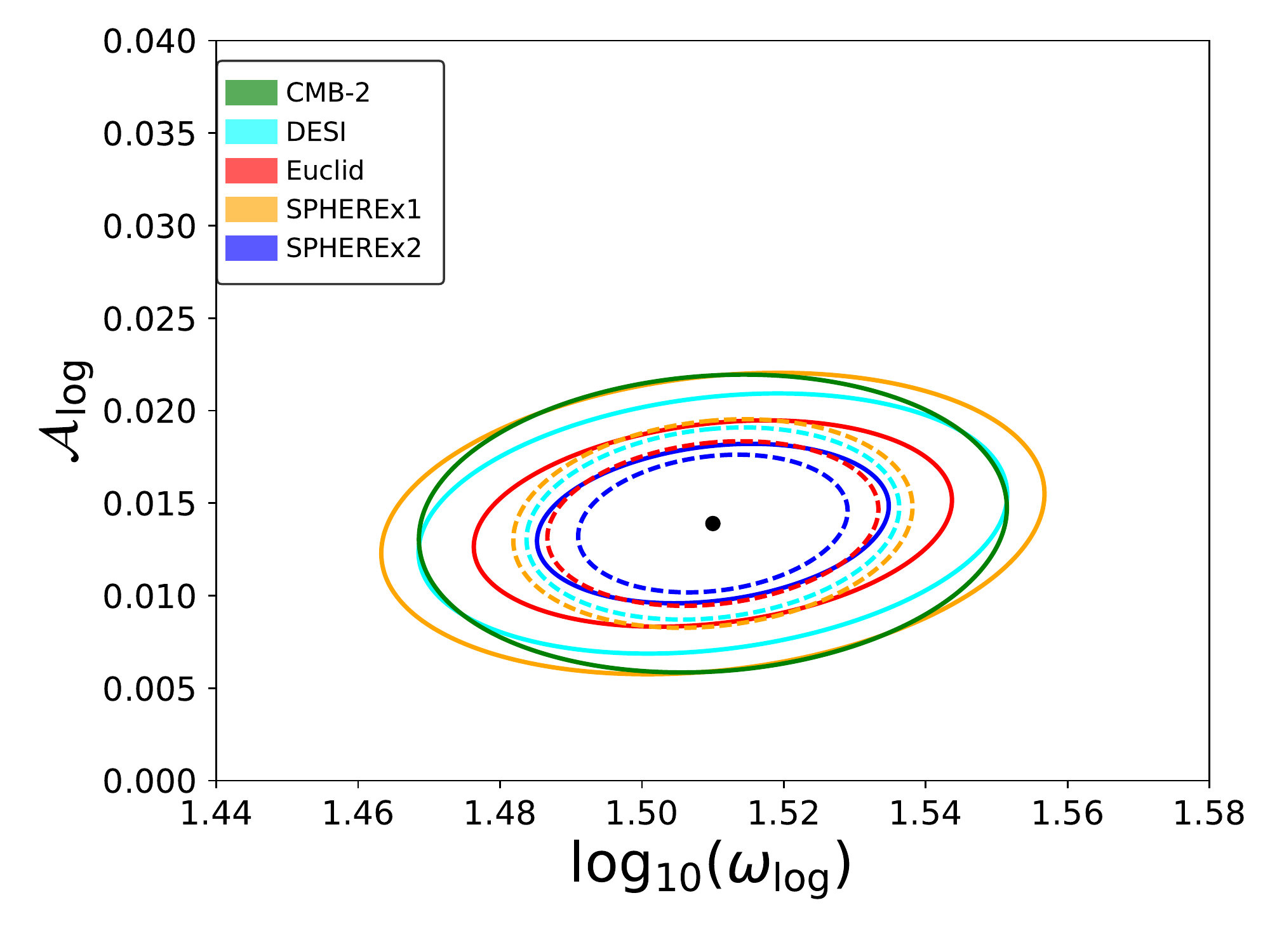}
\caption{Same plot as fig.~\ref{fig:wigg} but  with an amplitude smaller than by a factor 2.}
\label{fig:wigg2}
\end{figure}

\section{Conclusions}
\label{sec:conclusions}

In this paper we have studied the complementarity between the matter power spectrum from future 
galaxy surveys which have an accurate determination of redshift and cover a wide volume, such as DESI, 
Euclid and SPHEREx, and the one from the measurements of CMB anisotropies in temperature and 
polarization to help in characterizing primordial features in the PPS.
We have restricted ourselves to models predicting features which improve the fit to $Planck$ 2015 
temperature data with respect to the simplest power-law spectrum, although not at a statistical significant level.

By considering four representative deviations from a simple power-law PPS and including CMB 
uncertainties compatible with future measurements, we have shown that any of the surveys considered 
here with either a wide sky coverage and an accurate determination of redshift will be useful to 
decrease significantly the uncertainties in the features parameters, as is clear from 
figures~\ref{fig:cutoff}-\ref{fig:staro}-\ref{fig:step}-\ref{fig:wigg}-\ref{fig:wigg2}. 
As a best case from table~\ref{tab:sigma}, we have shown that 
the combination of information contained in the three surveys considered 
can detect the model super-imposed logarithmic oscillations at more than $3 \sigma$; 
we have explicitly checked that the same model with an amplitude smaller than by a factor 2 
($\mathcal{A}_{\rm log} = 0.0189$) can be detected at $3 \sigma$ by 
the combination of CMB and galaxy surveys considered here. 

The synergy with future galaxy surveys was also explored in previous works 
\cite{Gibelyouetal,Huangetal}. With respect to these works, our study has compared different 
galaxy surveys with the most updated specifications and has considered cosmological models which 
lead to an improved $\chi^2$ with respect to the simplest $\Lambda$CDM model, with the most recent 
data \cite{PlanckI2015,Aghanim:2015xee,PlanckXX2015}. Instead, previous works such as Gibelyou et 
al.~\cite{Gibelyouetal}, in which the model with an exponential cut-off was studied, and Huang et 
al.~\cite{Huangetal}, which considered the sharp edge in the first derivative of the potential, 
have adopted fiducial models with features in the PPS at comoving scales smaller than what current 
data seem to indicate. By choosing smaller comoving scales for the features, MI and MII could be 
more easily distinguished from a standard $\Lambda$CDM model, as the analysis of the fourth model 
explicitly shows.

Although not all (realistic and systematics) uncertainties have been taken into account in our 
forecasts, we have conservatively limited ourselves to the CMB angular power spectra of temperature 
and polarization fluctuations and to the power spectrum of galaxies from future surveys with an 
accurate determination of redshifts. We therefore believe that we can be optimistic even in probing 
features at large scales in the PPS for different reasons.
In the first instance, other surveys as LSST~\cite{Abell:2009aa} (photometric) and 
SKA~\cite{Maartens:2015mra} (radio) will access even larger volumes than the ones considered here. 
This aspect is particularly important since the features for three of the four models studied here 
seem effectively located at scales which are at the edge of those probed by DESI, Euclid and SPHEREx. 
Secondly, the deviations from a simple power law of primordial perturbations studied here can be 
accompanied by imprints in the CMB and/or galaxy shear, as well as in the CMB and/or galaxy 
bispectrum; these imprints in higher-order correlation functions can add to the ones we have 
considered here to further test primordial features. We hope to include some of these effects in our 
analysis elsewhere.
Finally, future CMB space missions \cite{Bouchet:2011ck,Kogut:2011xw,Matsumura:2013aja} will provide 
a final cosmic variance limited measurement of the E-mode polarization which will be crucial in 
discriminating a primordial origin of the features at $\ell \lesssim 40$ in the temperature power 
spectrum from a statistical fluctuation.

\section*{Acknowledgements}

We wish to thank Raul Abramo, Enzo Branchini, Xuelei Chen, Gigi Guzzo, Zhiqi Huang, Roy Maartens, 
Daniela Paoletti, Domenico Sapone and Emiliano Sefusatti for useful discussions and suggestions.  
We wish to thank Olivier Dor\'e and Roland de Putter for kindly providing the SPHEREx specifications.
We thank the anonymous referee for helpful comments and suggestions.
The support by the "ASI/INAF Agreement 2014-024-R.0 for the Planck LFI Activity of Phase E2" is 
acknowledged. We also acknowledge financial contribution from the agreement ASI n.I/023/12/0 
"Attivit\`a relative alla fase B2/C per la missione Euclid". LM acknowledges the grants MIUR PRIN 
2010-2011 "The dark Universe and the cosmic evolution of baryons: from current surveys to Euclid" 
and PRIN INAF 2012 "The Universe in the box: multiscale simulations of cosmic structure".
Preliminary results based on this work have been presented at the 28th Texas Symposium on Relativistic 
Astrophysics and at the Galileo Galilei Institute for Theoretical Physics during the workshop 
"Theoretical Cosmology in the Era of Large Surveys". We wish to thank the Kavli Institute for 
Theoretical Physics China where this work was partially carried out.

\section*{Note added}

After this work was nearly completed, a paper forecasting Euclid-like and LSST-like capabilities 
\cite{Chen:2016vvw} for models with features in the power spectrum appeared on the archive. 
Where a comparison is possible, our results are in qualitative agreement with those presented in 
\cite{Chen:2016vvw}.

\end{document}